\documentclass[aps,prd,superscriptaddress,showpacs,nofootinbib,amsfonts,amssymb,amsmath,notitlepage,balancelastpage,floatfix,10pt]{revtex4-1}

\usepackage[utf8x]{inputenc}
\usepackage[dvipdfmx]{graphicx}
\usepackage{wrapfig}
\usepackage{bm}
\usepackage[dvipdfmx]{color}
\usepackage{comment}
\usepackage{xcolor}
\usepackage[normalem]{ulem}
\usepackage{hyperref}
\usepackage{xspace}
\usepackage{xfrac}
\usepackage[nameinlink]{cleveref}
\usepackage{xifthen}

\hypersetup{
colorlinks=true,
citecolor=blue,
citebordercolor=red,
linktoc=all,
linkcolor=blue,
urlcolor=blue
}

\newcommand{\p}{\partial}
\newcommand{\al}[1]{\begin{align}#1\end{align}}
\newcommand{\nn}{\nonumber\\}
\newcommand{\df}{\text{d}}
\newcommand{\Tr}{{\rm Tr}\,}
\newcommand{\pmat}[1]{\begin{pmatrix}#1\end{pmatrix}}

\newcommand{\fn}[1]{\!\left(#1\right)}

\graphicspath{{./figs/}}
\newbox{\ORCIDicon}
\sbox{\ORCIDicon}{\large
                  \includegraphics[width=0.8em]{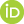}}

\begin{document}
\title{
Scaling solutions for gauge invariant flow equations in dilaton quantum gravity
}
%
\author{Yadikaer \surname{Maitiniyazi}\,\href{https://orcid.org/0009-0004-0826-1130}{\usebox{\ORCIDicon}}}
\email{ydqem22@mails.jlu.edu.cn}
\affiliation{Center for Theoretical Physics and College of Physics, Jilin University, Changchun 130012, China}
\author{Christof \surname{Wetterich}\,\href{https://orcid.org/0000-0002-2563-9826}{\usebox{\ORCIDicon}}}
\email{c.wetterich@thphys.uni-heidelberg.de}
\affiliation{Institut f\"ur Theoretische Physik, Universit\"at Heidelberg, Philosophenweg 16, 69120 Heidelberg, Germany}
\author{Masatoshi \surname{Yamada}\,\href{https://orcid.org/0000-0002-1013-8631}{\usebox{\ORCIDicon}}}
\email{m.yamada@kwansei.ac.jp}
\affiliation{Department of Physics and Astronomy, Kwansei Gakuin University,
Sanda, Hyogo, 669-1330, Japan}
\date{\today}

\begin{abstract}
We discuss the ultraviolet fixed point of asymptotically safe dilaton quantum gravity. It differs from the Reuter fixed point by the dependence of the Planck mass on a scalar field. The gauge invariant functional flow equation in the most general approximation with up to two derivatives strengthens the argument for the existence of this fixed point. The quantum effective action obtained from the scaling solution for dilaton quantum gravity can describe inflation for early cosmology and dynamical dark energy for late cosmology.
\end{abstract}

\maketitle

\section{Introduction}

Extensions of general relativity to models for the metric and a scalar singlet field play a key role in modern cosmology.
They are used for inflation \cite{Guth:1980zm,Sato:1981qmu,Linde:1983gd} and dynamical dark energy \cite{Wetterich:1987fm,Ratra:1987rm,Copeland:2006wr}.
The key ingredient is the quantum effective action for the metric coupled to the scalar.
The field equations from its variation of the effective action with respect to the fields are exact and include all quantum effects.
Cosmology is obtained as a solution of these field equations.
The issue is to find the correct quantum effective action $\Gamma$.
Employing diffeomorphism symmetry, one may use a derivative expansion with up to two derivatives
\al{
\label{eq:effectiveaction}
&\Gamma
=	\int \df^4x \sqrt{g}\left\{U(\rho) + \frac{K(\rho)}{2} g^{\mu\nu} \p_\mu{\phi}\p_\nu\phi -\frac{F(\rho)}{2}R \right\}\,.
}
For the scalar field $\phi$ we assume a symmetry $\phi\to -\phi$, with invariant 
\al{
\rho=\frac{1}{2}\phi^2 \,.
}
The curvature scalar $R$ involves two derivatives of the metric $g_{\mu\nu}$.
The effective action will also include higher derivative terms, for example $\sim R^2$.
Unless the coefficients of such terms are huge they do not play an important role for most cosmological epochs.

The effective action \labelcref{eq:effectiveaction} is the action of variable gravity \cite{Wetterich:2013jsa,Wetterich:2013aca,Wetterich:2013wza,Wetterich:2014gaa,Wetterich:2015ccd,WaliHossain:2014usl,Hossain:2014xha,Rubio:2017gty,Bettoni:2021qfs}, where the name reflects that the effective Planck mass $M_\text{p}$ can vary with $\phi$, $F(\rho)=M_\text{p}^2(\phi)$.
Variable gravity has been employed for many interesting cosmological models, in particular for dynamical dark energy, inflation, or quintessential inflation where both dynamical dark energy and inflation are described by the same scalar field $\phi$.
For a realistic model the effective action \labelcref{eq:effectiveaction} has to be supplemented by a part involving fields for the particles of the standard model.
Our quantum gravity computation will find that for large $\rho$ the function $F$ increases $\sim \phi^2$.
If also the particle masses increase $\sim \phi$ the ratio of particle mass over Planck mass remains constant even if $\phi$ varies with cosmological time.
This consequence of quantum scale symmetry \cite{Wetterich:2019qzx} makes variable gravity compatible with the strong observational constraints on time varying couplings or apparent violations of the equivalence principle.

The effective action of variable gravity involves three free functions of the scalar field---the effective potential $U(\rho)$, the kinetial $K(\rho)$ and the squared Planck mass $F(\rho)$.
Choosing them freely allows for a very wide variety of models.
Predictions for cosmology depend however, on the choice of these functions.
For a theory of quantum gravity one would like to predict these functions.
The approach of asymptotic safety \cite{Hawking:1979ig,Reuter:1996cp,Souma:1999at} is well suited for this purpose, since it can be based directly on a quantum field theory for the metric coupled to a scalar field, or more generally for gravitational fields coupled to fields for particles.
The modern developments of asymptotic safety for gravity are based on functional flow equations of the effective action \cite{Wetterich:1992yh}.
This approach introduces an infrared cutoff scale $k$ such that only fluctuations with momentum larger than $k$ are included.
Then $\Gamma_k$ becomes a $k$-dependent functional, and correspondingly $U$, $K$, and $F$ are functions of $\rho$ and $k$.
The functional flow equation describes the dependence of these objects on $k$.
The quantum effective action \labelcref{eq:effectiveaction} is obtained by solving the flow equations for $k\to 0$.

The key ingredient of asymptotic safety describing a nonperturbatively renormalizable theory of quantum gravity is the presence of an ultraviolet fixed point of the flow.
A solution valid for $k\to \infty$ permits the extension to arbitrarily short distances (ultraviolet completeness).
Such a fixed point has been found for  a large variety of truncations of the effective average action for pure gravity \cite{Lauscher:2002sq,Codello:2006in,%
Falls:2013bv,Falls:2014tra,Falls:2017lst,Falls:2018ylp,Kluth:2020bdv,Kluth:2022vnq,%
Dietz:2012ic,Dietz:2013sba,%
deBrito:2018jxt,Ohta:2013uca,Ohta:2015zwa,%
Benedetti:2009rx,Benedetti:2009gn,Benedetti:2010nr,Groh:2011vn,%
Manrique:2010am,%
Donkin:2012ud,%
Christiansen:2012rx,Christiansen:2014raa,Christiansen:2016sjn,Christiansen:2017bsy,%
Eichhorn:2018akn,Eichhorn:2018ydy,%
Codello:2013fpa,Demmel:2014hla,Biemans:2016rvp,Gies:2016con,deBrito:2020rwu,deBrito:2020xhy,deBrito:2021pmw,%
Gonzalez-Martin:2017gza,Baldazzi:2021orb,Baldazzi:2021fye,deBrito:2022vbr,Pawlowski:2020qer,Falls:2020qhj,Knorr:2021slg,%
Mitchell:2021qjr,Morris:2022btf,Sen:2021ffc,deBrito:2022vbr,Eichhorn:2022ngh,Saueressig:2023tfy,Becker:2024tuw,Kawai:2024rau,Kawai:2025wkp,Saueressig:2025ypi,Pawlowski:2025etp} and for gravity-matter systems \cite{Dona:2013qba,Dona:2015tnf,Percacci:2015wwa,Oda:2015sma,Hamada:2017rvn,Meibohm:2015twa,%
Labus:2015ska,%
Eichhorn:2016esv,Eichhorn:2015bna,%
Christiansen:2017cxa,Meibohm:2016mkp,%
Biemans:2017zca,%
deBrito:2019epw,Eichhorn:2018nda,%
Alkofer:2018fxj,Alkofer:2018baq,%
Burger:2019upn,%
deBrito:2019umw,deBrito:2020dta,Eichhorn:2020kca,Eichhorn:2020sbo,Eichhorn:2021tsx,%
Ohta:2021bkc,Laporte:2021kyp,Knorr:2022ilz,Hamada:2020mug,Eichhorn:2022vgp,Wetterich:2022bha,%
Pastor-Gutierrez:2022nki,Eichhorn:2021qet,Eichhorn:2022vgp,Eichhorn:2022gku,Eichhorn:2023gat,Pawlowski:2023gym,Eichhorn:2023xee,deBrito:2023kow,deBrito:2023ydd,deBrito:2023pli,Korver:2024sam,Eichhorn:2025ezh,deBrito:2025nog,Kher:2025rve,Riabokon:2025ozw,Ohta:2025xxo,Bonanno:2025qsc}.
It allows predictions for parameters of the standard model \cite{Shaposhnikov:2009pv,Eichhorn:2017eht,Eichhorn:2017ylw,Eichhorn:2018whv,Alkofer:2020vtb,Eichhorn:2017muy,Pawlowski:2018ixd,Harst:2011zx,Christiansen:2017gtg,Eichhorn:2017lry,Pastor-Gutierrez:2024sbt,Eichhorn:2025sux,Eichhorn:2025ilu} and its extensions~\cite{Eichhorn:2019dhg,Eichhorn:2017als,Reichert:2019car,Wetterich:2019zdo,Hamada:2020vnf,%
 Kowalska:2020zve,Kowalska:2020gie,Kowalska:2022ypk,Chikkaballi:2022urc,%
 Boos:2022jvc,Boos:2022pyq,deBrito:2021akp,deBrito:2025ges,Assant:2025gto}.
In our approximation \labelcref{eq:effectiveaction} an ultraviolet fixed point requires a scaling solution for which $U/k^4$, $F/k^2$, and $K$ become functions of a dimensionless scaling field without any explicit dependence on $k$.
One can then start the flow at arbitrary short distances or arbitrarily large $k$ and follow it to macroscopic distances or small $k$.

Most of the previous work on asymptotic safety concentrates on the Reuter fixed point and its extension to matter, with $\rho$-independent $U$, $K$, $F$ for the scaling solution.
For our purpose we are interested in a different fixed point, for which the scaling solution is characterized by $F\sim \rho$ for $\rho\to \infty$.
This is named dilaton quantum gravity \cite{Henz:2013oxa,Henz:2016aoh}, since for field-independent $F/\rho$, $K$, and vanishing $U$, the effective action \labelcref{eq:effectiveaction} is invariant under dilatation or scale transformations.
For dilaton quantum gravity this holds for $k\to 0$ or $\rho\to \infty$ provided $U$ vanishes $\sim k^4$.
For nonzero $\phi$ the dilatation symmetry is spontaneously broken, producing the dilaton as a Goldstone boson.
The aim of the present paper is to find scaling solutions for the approximation \labelcref{eq:effectiveaction} for which $U$, $F$, and $K$ are nontrivial functions of $\rho$.
For the scaling solution of dilaton quantum gravity $U$ has been found previously to be almost independent of $\rho$, while $F$ approaches a constant for $\rho\to 0$ and is dominated by the dimensionless ``nonminimal" coupling $\xi$ for large $\rho$, corresponding to a term $\sim \xi \phi^2 R$ in the effective action.

While these qualitative features appear to be rather robust, a reliable computation of the kinetial $K(\rho)$ has to face certain challenges.
The first concerns the fact that for $F=\xi\phi^2=2\xi\rho$ with constant $\xi$ the kinetial can be negative.
For constant $K$ and $\xi$ stability requires $K/\xi>-6$.
This is due to a mixing in the kinetic terms for the scalar ``cosmon"-field and the metric.
Stability is most easily seen by a field transformation to the Einstein frame, where the coefficient of the scalar kinetic term is $\sim K+6\xi$.
A reliable computation should therefore cover also the range of negative $K$ as long as stability is preserved.
Second, for constants $F/\rho$ and $K=-3F/\rho$ with $U=0$ the effective action admits an enhanced symmetry, namely conformal symmetry and local Weyl symmetry.
In this limit the scalar field is no longer a dynamical degree of freedom.
One would like a computation of $K(\rho)$ which respects this symmetry.
Third, for constant $U$ and $F$ and constant $K\rho$ the effective action is invariant under a multiplicative rescaling of $\phi$, while keeping the metric invariant.
Again, one would like to find a version of the renormalization flow that respects this additional symmetry.

The consistency with diffeomorphism symmetry, together with the additional symmetries mentioned above, poses a technical problem for the functional flow equations.
If one wants to preserve these symmetries, the infrared cutoff term involves the macroscopic fields $g_{\mu\nu}$ and $\phi$, for example by the appearance of covariant derivatives.
We base our computation on the setting of a gauge invariant flow equation \cite{Wetterich:2016ewc}.
The field dependence of the cutoff introduces correction terms to the simple one-loop form of the exact functional flow equation.
In the setting of the simplified flow equation \cite{Wetterich:2024ivi} these corrections have been estimated and found to be small.
We omit them in the present work.
The gauge invariant flow equation is, in this approximation, identical to the background field method with a particular ``physical gauge fixing," which removes the gauge modes in the metric without affecting the ``physical fluctuations"~\cite{Pawlowski:2018ixd}.

\section{Flow equation and scaling equation}

Our starting point is the flow equation for the effective average action \cite{Wetterich:1992yh}
\al{
\label{eq:floweq}
\p_t \Gamma_k =\frac{1}{2}\Tr \left[ \left( \Gamma_k^{(2)} +\mathcal R_k \right)^{-1}\p_t \mathcal R_k\right]\,,
}
where $\p_t =k \p_k$.
Here $\mathcal R_k$ is the infrared cutoff, $\Gamma_k^{(2)}$ is the second functional derivatives of $\Gamma_k$, and $(\Gamma_k^{(2)}+\mathcal R_k)^{-1}$ is the propagator in the presence of fields and the infrared cutoff.
For a discussion of the properties of the flow equation and its application in many fields we refer to Refs.~\cite{Reuter:1993kw,Morris:1998da,Berges:2000ew,Aoki:2000wm,Bagnuls:2000ae,%
  Polonyi:2001se,Pawlowski:2005xe,Gies:2006wv,Delamotte:2007pf,Sonoda:2007av,Igarashi:2009tj,%
  Rosten:2010vm,Braun:2011pp,Dupuis:2020fhh}.
We define dimensionless quantities by
\al{
&u(\tilde\rho) = U(\rho)/k^4\,,&
&w(\tilde\rho) = F(\rho)/2k^2\,,&
&\kappa(\tilde\rho) = K(\rho)\,,&
&\tilde\rho= \rho/k^2\,.
}
The flow equation for $u$ can be obtained by evaluating \cref{eq:floweq} for a constant scalar field and flat metric $g_{\mu\nu}=\eta_{\mu\nu}$.
For $w$ and $\kappa$, one adds small inhomogeneities in the macroscopic field $\phi$ and $g_{\mu\nu}$, for which we evaluate the flow equation by the use of the heat kernel method.
One obtains
\al{
\p_t u&=\beta_U
=-\p_\tau u-4u+\frac{1}{32\pi^2}M_U\,,
\label{eq: beta function of U}
\\[1ex]
\p_t w&=\beta_F\
=-\p_\tau w-2w - \frac{1}{96\pi^2}M_F\,,
\label{eq: beta function of W}
\\[1ex]
\p_t \kappa&=\beta_K
=-\p_\tau \kappa + M_K\,,
\label{eq: beta function of K}
}
where $M_U$, $M_F$, and $M_F$ are the flow kernels.
The computation of $M_U$, $M_F$, and $M_K$ is a main technical achievement of this work. 
We present this in various appendices.
The switch from fixed $\phi$ to fixed $\phi/k$ or fixed $\tilde\rho$ introduces terms involving
\al{
-\tilde\phi\, \p_{\tilde\phi}= -2{\tilde \rho}\,\p_{\tilde \rho} = \p_\tau \,,
}
where $\tau$ is the logarithm of the dimensionless field
\al{
\tau=-\log \sqrt{2\tilde\rho}=-\log |\tilde \phi| = -\log |\phi/k|  \,.
\label{eq: scales tau rho}
}
The terms $-4u$ and $-2w$ reflect the dimension of $U$ and $F$.

A possible scaling solution is obtained as a solution of \cref{eq: beta function of U,eq: beta function of W,eq: beta function of K} for which $u$, $w$, and $\kappa$ only depend on $\tilde\rho$ without any explicit dependence on $k$.
This generalizes the notion of a fixed point for a finite number of couplings.
The functions $u$, $w$, and $\kappa$ encode an infinite number of couplings, defined, for example, as the coefficients of an expansion around $\tilde\rho=0$.
For the scaling solution, all these couplings take fixed values.
We focus on this type of scaling solution.
Further scaling solutions may be obtained by switching to renormalized dimensionless scaling fields reflecting the presence of an anomalous dimension.
Such fields are related to $\tilde\rho$ by a nonlinear field transformation, see below.

The scaling solution for the functions $u(\tilde\rho,k)$ etc. requires 
\al{
\p_t u=\p_t w=\p_t \kappa=0\,.
}
It has therefore to obey the ``scaling equations"
\al{
&\p_\tau u(\tilde\rho)=-4u(\tilde\rho)+\frac{1}{32\pi^2}M_U\,,
\label{eq: u full}
\\[1ex]
&\p_\tau w(\tilde\rho)=-2w(\tilde\rho) - \frac{1}{96\pi^2}M_F\,,
\label{eq: w full}
\\[1ex]
&\p_\tau \kappa(\tilde\rho)=M_K\,,
\label{eq: K full}
}
This is a coupled system of three nonlinear differential equations.
Its form is very similar to the flow equations for the dependence of $k$ for the couplings $u_0=u(\tilde\rho=0)$, $w_0=w(\tilde\rho=0)$, $\kappa_0=\kappa(\tilde\rho=0)$.
The scaling equations determine, however, three functions of $\tilde\rho$, and not only three couplings.
This is of key importance for their possible solutions.
They have to be defined for the whole range $0\leq \tilde\rho < \infty$.
As is common for differential equations the corresponding boundary conditions pose severe constraints.
The flow kernels $M_U$, $M_F$, $M_K$ in \cref{eq: beta function of U,eq: beta function of W,eq: beta function of K} involve first and second derivatives of $\tilde\rho$.
The differential equations are second order.
The general local solution therefore depends on six initial values $u(\tilde\rho_0)$, $\p_{\tilde\rho}u(\tilde\rho_0)$, $w(\tilde\rho_0)$, $\p_{\tilde\rho}w(\tilde\rho_0)$, $\kappa(\tilde\rho_0)$, $\p_{\tilde\rho}\kappa(\tilde\rho_0)$ taken at some initial values $\tilde\rho_0$.
These initial values define a family of local scaling solutions.
The boundary conditions for a global solution for all $\tilde\rho$ selects a subset of this family.

\section{Scaling solution}

\begin{figure}
\centering
\includegraphics[width=0.48\columnwidth]{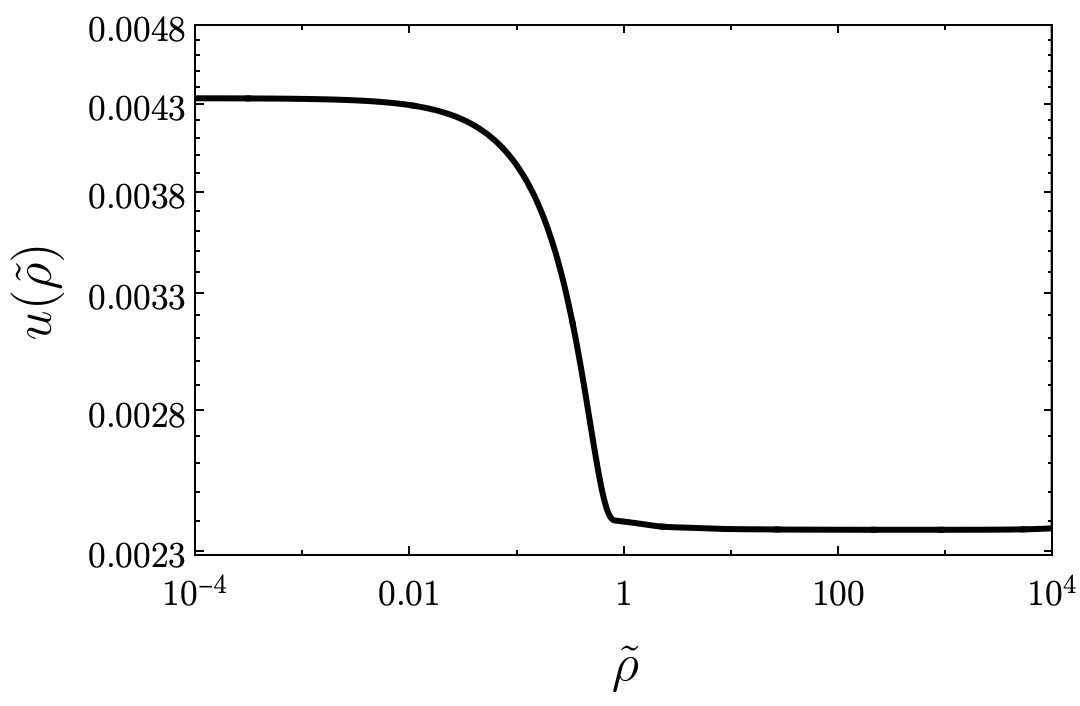}
    \hspace{2ex}
\includegraphics[width=0.48\columnwidth]{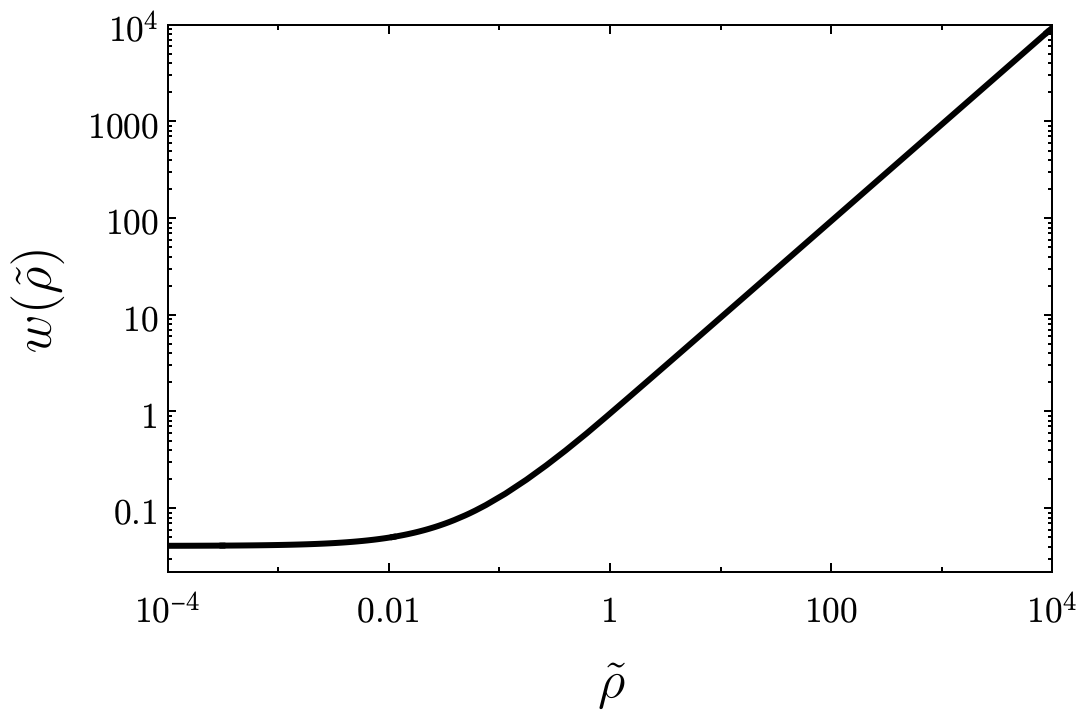}
\includegraphics[width=0.48\columnwidth]{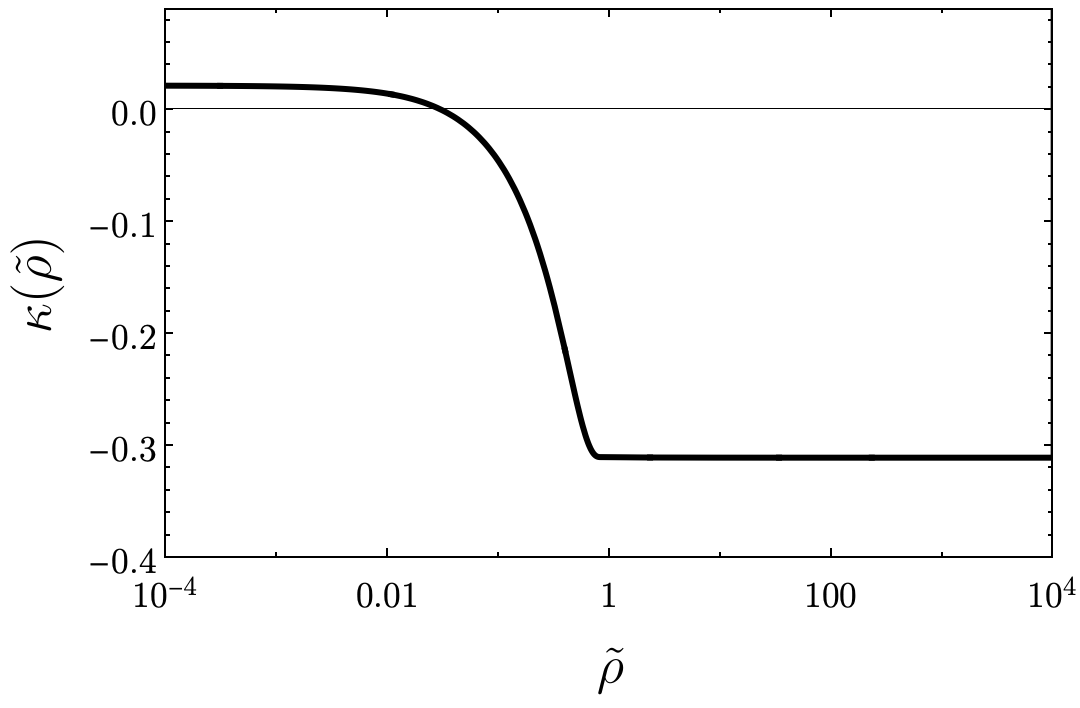}
    \hspace{2ex}
\includegraphics[width=0.46\columnwidth]{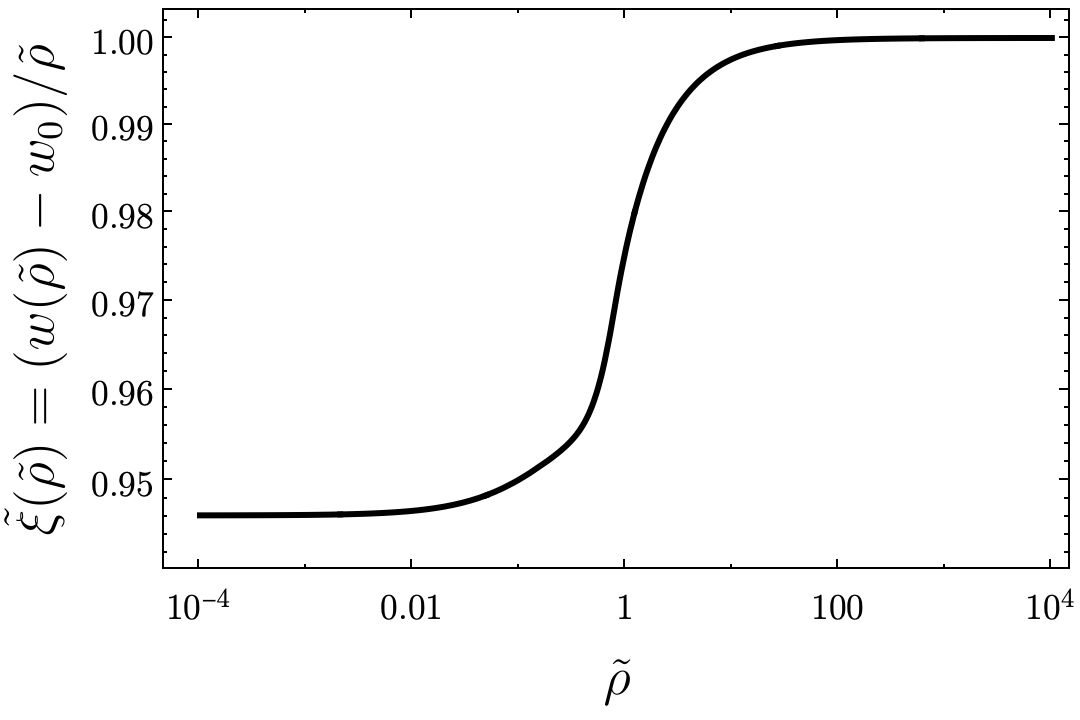}
    \caption{
    Scaling solution for $u(\tilde\rho)$ (top-left), $w(\tilde\rho)$ (top-right), and $\kappa(\tilde\rho)$ (bottom-left). The initial condition is set at $\tilde\rho_0=10^4$ to be $u(\tilde\rho_0)=u_\infty =3/(128\pi^2)\approx 0.00238$, $\p_{\tilde\rho}u(\tilde\rho_0)\approx 0$, $w(\tilde\rho_0)= w_0 + \xi_\infty\tilde\rho_0$, $\p_{\tilde\rho} w(\tilde\rho_0)= \xi_\infty$ with $w_0=0.0277$, $\xi_\infty=1$, $\kappa(\tilde\rho_0)=\kappa_\infty=-0.3$, and $\p_{\tilde\rho}\kappa(\tilde\rho_0)\approx 0$. This initial condition corresponds to $B=6+\varepsilon_\infty =6 + \kappa_\infty/\xi_\infty =5.7$.
    The scaling solution for $\tilde\xi(\tilde\rho)=(w(\tilde\rho)-w_0)/\tilde\rho$ is displayed in the bottom-right panel. For large field value $\tilde\rho\to \infty$, this function converges to $\xi_\infty=1$.
    }
    \label{fig:u and w and k scaling solution}
\end{figure}
A numerical solution of the scaling equations which extends to the whole range of $\tilde\rho$ is shown in \Cref{fig:u and w and k scaling solution}.
We observe simple characteristic features.
The function $u(\tilde\rho)$ interpolates between two constants $u_0=u(\tilde\rho=0)$ and $u_\infty=u(\tilde\rho\to \infty)$.
We will see that $u_\infty$ is simply determined by the number of massless physical excitations, two for the graviton and one for scalar.
For $u_0$ the gravitational contribution is somewhat modified which leads to a small shift.
The function $w(\tilde\rho)$ increases linearly for large $\tilde\rho$, reflecting the nonminimal gravitational coupling of the scalar.
It approaches for $\tilde\rho\to \infty$ the form $w(\tilde\rho)\to \infty = \xi_\infty \tilde\rho$ with constant $\xi_\infty$.
The approximated form $w(\tilde\rho)=w_0+ \xi_\infty \tilde\rho$ shows a crossover for $\tilde\rho_\text{c}=w_0/\xi_\infty$, with $w_0=w(\tilde\rho=0)$.
This value determines the location of the crossover for the potential $u(\tilde\rho)$.
We see in \Cref{fig:u and w and k scaling solution} that this form is a good approximation by plotting $\tilde\xi(\tilde\rho)=(w(\tilde\rho) - w_0)/\tilde\rho$, which corresponds to $w(\tilde\rho)=w_0+\tilde\xi(\tilde\rho) \tilde\rho$.
The function $\xi(\tilde\rho)$ varies only mildly between $\tilde\rho=0$ and $\tilde\rho\to \infty$.
Finally, the kinetial $\kappa(\tilde\rho)$ also shows a crossover between two plateaus $\kappa_0 = \kappa(\tilde\rho\to 0)$ and $\kappa_\infty=\kappa(\tilde\rho\to \infty)$, located near $\tilde\rho_c$.
For the solution shown in \Cref{fig:u and w and k scaling solution} one has $\kappa_\infty<0$, demonstrating our possibility to cover the range of negative $K$.

We can associate the asymptotic behavior with two fixed points.
The infrared (IR) fixed point is approached for $\tilde\rho\to \infty$, or $k\to 0$ at fixed $\phi$.
The ultraviolet (UV) fixed point characterizes the limit $\tilde\rho\to 0$, or $k\to \infty$ at fixed $\phi$.
The scaling solution describes a crossover between the two fixed points as a function of $\tilde\rho$.
The discussion of the UV completion of quantum gravity often considers only the couplings for $\tilde\rho\to 0$, e.g., $u_0$, $w_0$ and $\kappa_0$.
The functional form of the flow equations implies, however, that fixed points for the $k$ flow of these couplings only exist if the infinitely many couplings encoded in the functions $u(\tilde\rho)$, $w(\tilde\rho)$, $\kappa(\tilde\rho)$ also take fixed values.
The UV completion of quantum gravity requires a scaling solution for these functions, which imposes constraints for the existence of the scaling solution for $\tilde\rho\to \infty$.
This provides a deep connection between the IR and the UV.
Quantum gravity becomes relevant for the behavior of the effective potential $U(\rho)$ for large $\rho$, which typically determines the late time behavior of dynamical dark energy within variable gravity models.

\begin{figure}
    \centering
    \includegraphics[width=0.48\columnwidth]{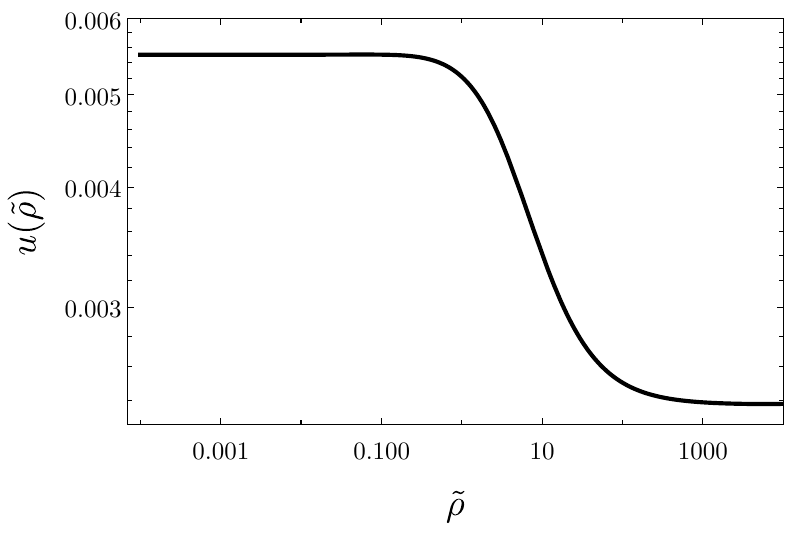}
    \hspace{2ex}
        \includegraphics[width=0.48\columnwidth]{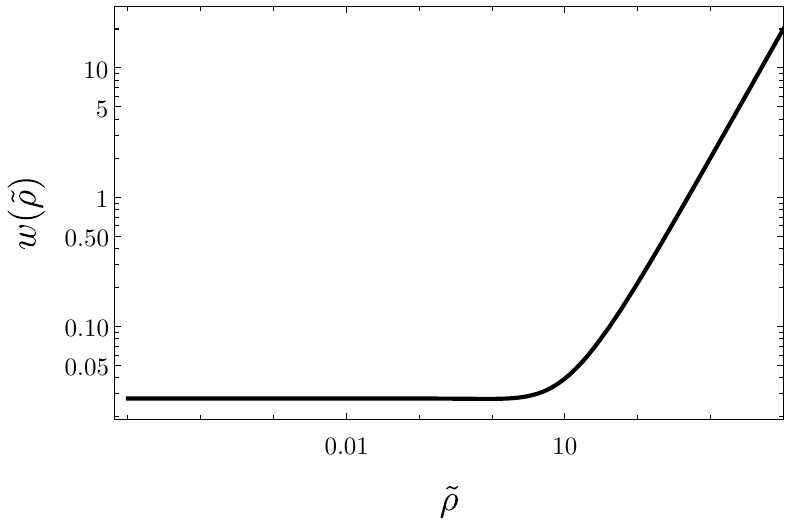}
    \caption{
    Scaling solution for $u(\tilde\rho)$ (left) and $w(\tilde\rho)$ (right) with $\kappa(\tilde\rho)=1$ fixed for all field values. 
The initial condition is set at $\tilde\rho_0=10^4$ to be $u(\tilde\rho_0)=u_\infty =3/(128\pi^2)\approx 0.00238$, $\p_{\tilde\rho}u(\tilde\rho_0)\approx 0$, $w(\tilde\rho_0)=w_0 + \xi_\infty \tilde\rho_0$ and $\p_{\tilde\rho}w(\tilde\rho_0)=\xi_\infty$ with $w_0=0.0277$ and $\xi_\infty=0.07$.
    }
    \label{fig:u and w scaling solution}
\end{figure}

The qualitative behavior of the crossover for the functions $u(\tilde\rho)$ and $w(\tilde\rho)$ is rather robust and does not depend strongly on the behavior of $\kappa(\tilde\rho)$.
This may be demonstrated by solving the flow equations, \labelcref{eq: beta function of U,eq: beta function of W}, for a fixed value $\kappa(\tilde \rho)=1$.
The result is shown in \Cref{fig:u and w scaling solution}.
The initial values used here are $u(10^{4})=u_\infty=3/(128\pi^2)=0.00237975$, $\p_{\tilde\rho}u(10^{4})\approx 0$, $w(10^{4})=w_0 + \xi_\infty \tilde\rho_0=19.39$, and $\p_{\tilde\rho}w(10^{4})=\xi_\infty=0.07$ as a boundary condition.

\section{Infrared and ultraviolet fixed points}

The IR and UV-fixed point behavior for $\tilde\rho\to \infty$ and $\tilde\rho\to 0$ is seen clearly in \Cref{fig:u and w and k scaling solution,fig:u and w scaling solution}.
It can be understood quantitatively.
For the limits $\tilde\rho\to \infty$ and $\tilde\rho\to 0$ the flow generators can be expanded in powers of $\tilde\rho^{-1}$ or $\tilde\rho$, respectively.
Keeping only few terms, we can write down explicit expressions.
This permits an analytical treatment of the IR and UV-fixed points.

\subsection{Large field expansion and IR fixed point}

The properties of the IR fixed point are easy to understand.
With dimensionless Planck mass $\sqrt{2\xi_\infty \tilde\rho}$ going to infinity, the gravitational coupling approaches zero.
Thus, the metric fluctuations decouple, except for a constant contribution $M^\text{gr}_U=2$, reflecting the 2 degrees of freedom of the massless graviton.
For constant $u(\tilde\rho\to \infty)=u_\infty$, the scalar potential is flat and we are left with a massless scalar field.
For $\kappa(\tilde\rho\to\infty)=\kappa_\infty$ there are no interactions from the kinetic term either.
We are then left with a decoupled system of a free massless scalar and a massless graviton.
The contribution of the massless scalar to the flow of the cosmological constant $U(\tilde\rho\to \infty)$ is given by $M_U^{\text{s}}=1$.
The scaling solution for $u(\tilde\rho\to \infty)$ is given by $u=M_U/(128\pi^2)$, resulting in 
\al{
u_\infty = \frac{3}{128\pi^2} \approx 0.00238\,.
}
This is the value which is approached in \Cref{fig:u and w and k scaling solution,fig:u and w scaling solution}.
For the scaling solution the cosmological constant, $U=u_\infty k^4$ vanishes for $k\to 0$.
Since gravity effectively decouples the nonminimal coupling $\xi$ and the kinetial $\kappa$ become independent of $\tilde\rho$.
For the IR fixed point they are free parameters.
Since physics is not affected by a change of the normalization of the scalar field, only the ratio $\varepsilon_\infty=\kappa_\infty/\xi_\infty$ matters.
We end with a continuous family of IR fixed points parametrized by $\varepsilon_\infty$.
Stability requires $\varepsilon_\infty>-6$.
Since $\kappa(\tilde\rho)$ becomes a constant $k\p_k K$ also becomes a constant.
There is therefore no anomalous dimension for the scalar field, implying that $\phi/k$ is the appropriate dimensionless renormalized field for the IR fixed point.

For finite large $\tilde \rho$ the gravitational effects start to matter, being suppressed by powers of $\tilde\rho^{-1}$.
The scalar field and the metric are no longer decoupled.
In this large field region we can perform a series expansion.
To this end, we define $y=1/\tilde\rho$ and make the expansion ansatz
\al{
\label{eq:uyexpansion}
u(y) &= u_\infty + \sum_{n=1}^\infty  \frac{s_n}{n!\,\xi_\infty^n} y^n= u_\infty + \frac{s_1}{\xi_\infty}y + \cdots\,, \\
\label{eq:wyexpansion}
w(y) &= \xi_\infty y^{-1} + \sum_{n=0}^\infty \frac{t_n}{n! \, \xi_\infty^n} y^n = \xi_\infty y^{-1} + t_0 + \frac{t_1}{\xi_\infty} y+ \cdots\,, \\
\label{eq:kyexpansion}
\kappa(y) &= \varepsilon_\infty \xi_\infty + \sum_{n=1}^\infty  \frac{k_n}{n! \, \xi_\infty^{n-1}} y^n = \varepsilon_\infty \xi_\infty  + k_1 y + \cdots\,,
}
where $\varepsilon_\infty=\kappa_\infty/\xi_\infty$.
Switching to the variable $x =y/\xi_\infty$ we expand the scaling solution in terms of $x$
\al{
\label{eq:uexpansion}
\p_\tau u(x)&=2 x\p_x u(x)= -4u(x)+ A_u +\frac{C_u}{B}x = 2s_1x +\cdots\,,
\\[1ex]
\label{eq:wexpansion}
\p_\tau w(x)&=2 x\p_x w(x)= -2w(x) +\frac{A_w}{B}+\frac{C_w}{B^2}x+\cdots = -\frac{2}{x} + 2t_1 x +\cdots\,,
\\[1ex]
\label{eq:kexpansion}
\p_\tau \kappa(x) &=2 x\p_x \kappa(x) = \xi_\infty \frac{C_\kappa}{B^2}x +\cdots = 2\xi_\infty k_1 x +\cdots\,.
}
Here $B=6+\varepsilon_\infty$ has to be positive for stability.
For the conformal point $B=0$ the expansion has to be modified.
The coefficients of the IR expansion \labelcref{eq:uexpansion,eq:wexpansion,eq:kexpansion} read
\al{
A_u &=\frac{3}{32\pi^2}\,,
\\[1ex]
C_u &=\frac{1}{480\pi^2} (34 B t_0+75 B u_\infty -54t_0 + 5 k_1)\,,
\\[1ex]
A_w &=\frac{1}{1536 \pi ^2 }(509 B+194),
\\[1ex]
C_w &=\frac{1}{322560\pi^2}(37950 B^2 t_0+85505 B^2 u_\infty -20700 B t_0-1680 B k_1+5124 B u_\infty +245808 t_0-41496 k_1-69804 u_\infty)\,,
\\[1ex]
C_\kappa &=-\frac{9}{16 \pi ^2}+\frac{21B}{32 \pi ^2}-\frac{5 B^2}{64 \pi ^2}-\frac{B^3}{384 \pi ^2}\,.
}
The terms with factor $B^{-1}$ involve the scalar fluctuations.
There are no poles or other nonanalytic features as long as the stability condition $B>0$ is met.
This shows that our method can cover the whole stable range for $K$, including negative values.

The fixed point values for the coefficients $t_0$, $s_1$, $t_1$, $k_1$ are obtained from \cref{eq:uexpansion,eq:wexpansion,eq:kexpansion} by equating the coefficients of $x^n$ on both sides
\al{
&t_0 = \frac{A_w}{2B} = \frac{509}{3072\pi^2}+ \frac{97}{1536\pi^2 B}\,.
}
Also $k_1$ depends on $B$
\al{
k_1=\frac{C_\kappa}{2B}=\frac{21}{64 \pi ^2B}-\frac{9}{32 \pi ^2 B^2}-\frac{5 }{128 \pi ^2}-\frac{B}{768 \pi ^2}\,.
}
For $s_1$ and $t_1$ we can then solve iteratively
\al{
&s_1 = \frac{C_u}{6B}\,,&
&t_1 = \frac{C_w}{4B^2}\,.
}
These coefficients diverge for $B\to 0$, indicating that in the close vicinity of the conformal point $B=0$, a different approximation may be more suited.
For $B=8$~($B=50$) we find $A_w=0.281~(1.692)$, $C_u=0.046t_0 + 0.00106 k_1 + 0.127 u_{\infty}~(0.347 t_0+0.00106k_1 + 0.792u_{\infty})$, and $C_\kappa=-0.167~(-49.50)$.
This can be compared with the values in Ref.~\cite{Henz:2016aoh}, namely, $A_w=-4.419 ~(-1025.99)$, $C_u= -5.0323t_0 - 2.516k_1 + 46.493 u_{\infty}~(
3244.03 t_0 + 73.728 k_1 + 77139.1 u_{\infty})$, and $C_\kappa=-3.49~(76629.8)$.
A more detailed comparison can be found in \Cref{sec:Comparison}.
While the overall features of the dilaton quantum gravity fixed point are rather robust, the quantitative details depend on the particular version of the functional flow equation.

We can use the expansion \labelcref{eq:uyexpansion,eq:wyexpansion,eq:kyexpansion} in order to set initial conditions for  numerical solutions of the scaling equations,
We take $x$ instead of $\tilde\rho$ and set initial conditions for some $x_\text{in} \ll 1$.
With $u(x)=u_\infty + s_1x$, fixed $u_\infty$, and $s_1$ a function of $B$, the initial conditions $u(x_\text{in})$ and $\p_x u_\text{in}$ are fixed as functions of $B$.
For $w=\frac{1}{x} + t_0 + t_1 x$ and $t_0$, $t_1$ functions of $B$, $w(x_\text{in})$ and $\p_x w(x_\text{in})$ are also fixed functions of $B$.
If we use $\varepsilon(x) = \kappa(x)/\xi_\infty=B-6 + k_1 x$, the initial conditions for $\varepsilon(x_\text{in})$ and $\p_x \varepsilon(x_\text{in})$ are determined as functions of $B$ as well. 
We end with a one-parameter family of local scaling solutions parametrized by $B$.
This feature persists if one continues the series to higher power of $x$.
Formally, a second free parameter is $\xi_\infty$.
This is, however, only fixing the overall normalization of the scalar field.
Without loss of generality one may use a normalization defined by $\xi_\infty=1$.
In this case the scalar field $\phi$ denotes precisely the Planck mass in the IR limit.

\subsection{Small field expansion and UV-fixed point}

With initial conditions for some small $x_\text{in}$ fixed by $B$, the scaling solution has to be continued to the UV-fixed point for $\tilde\rho\to 0$.
The existence of such a global solution may select a specific value of $B$.
The scaling solution has to exist for the whole range of an appropriate dimensionless ``scaling field". 
{\it A priori}, it is not obvious which is the form of the scaling field in the UV region.
It may be a renormalized field in the case of a nonzero anomalous dimension.
Because of the simple properties of the IR fixed point, which describes a free theory, the anomalous dimension vanishes in the IR limit, justifying the choice of $\phi/k$ as a dimensionless renormalized scaling field.
The UV limit may be characterized by an interacting theory.
An anomalous dimension is therefore possible.
For example, if $K$ diverges proportional $\phi^{-2}$ for $\phi\to 0$, the renormalized scaling field is $\phi$ instead of $\phi/k$.
In this case the anomalous dimension cancels the canonical dimension of the field.
In the present paper we concentrate on scaling solutions for which the anomalous dimension of the scalar field vanishes in the UV.
This is the case for solutions for which $\kappa(\tilde\rho\to 0)$ reaches a constant $\kappa_0$, as in \Cref{fig:u and w and k scaling solution}.
The scaling solution is then a solution of the system of differential equations \labelcref{eq: u full,eq: w full,eq: K full} which exists for the whole range of $\tilde\rho$.

Near the UV-fixed point we approximate the scaling solution by an expansion in power of $\tilde\rho$.
A first approximation inserts on the rhs of the scaling equation the ansatz
\al{
\label{eq:UVansatz}
&u(\tilde\rho) = u_0\,,&
&w(\tilde\rho) = w_0\,,&
&\kappa(\tilde\rho) = \kappa_0\,.
}
If one can find a solution with this ansatz it would correspond to a full solution for all $\tilde\rho$ in our truncation, corresponding to the extended Reuter fixed point.
The effective action would be invariant under the shift $\phi\to \phi +c$.
Insertion of the ansatz \labelcref{eq:UVansatz} in the scaling equation yields
\al{
  \label{eq:uExpansionSmall2}
&\p_\tau u(\tilde\rho)=
  -4u_0 -\frac{5 u_0}{24 \pi ^2 (u_0-w_0)} -\frac{3 u_0}{160 \pi ^2 (u_0-4 w_0)} +\frac{79}{480 \pi ^2}\,,
  \\[1ex]
  \label{eq:wExpansionSmall2}
&\p_\tau w(\tilde\rho)=
 -2w_0-\frac{25 u_0}{64 \pi ^2 (u_0-w_0)} + \frac{5 u_0^2}{1344 \pi ^2 (u_0-4 w_0)^2} -\frac{9 u_0}{896 \pi ^2 (u_0-4 w_0)} +  \frac{1207}{2688 \pi ^2}\,,
\\[1ex]
\label{eq:kappaExpansionSmall2}
&\p_\tau\kappa(\tilde\rho) =
\frac{ (5 u_0-23 w_0)}{240 \pi ^2 (u_0-4 w_0)^2}\kappa_0\,.
}
The lhs of these equations vanishes for $\tilde\rho=0$, leaving us with three equations for $u_0$, $w_0$ and $\kappa_0$.
The first two equations do not involve $\kappa_0$ and fix
\al{
&u_{0*}=0.0055\,,&
&w_{0*}=0.0277\,.
}
For these values \cref{eq:kappaExpansionSmall2} has no solution for $\kappa_0>0$.
Requiring $\kappa_0>0$ as a condition for stability the system \labelcref{eq:uExpansionSmall2,eq:wExpansionSmall2,eq:kappaExpansionSmall2} is overdetermined.
A constant solution \labelcref{eq:UVansatz} does not exist.

One concludes that for solutions with a vanishing anomalous dimension, the higher order terms in the expansion $u(\tilde\rho)=u_0 + c_1 \tilde\rho+\cdots$ etc. have to contribute to the scaling equation for $\tilde\rho=0$.
This implies that a scaling solution with $\tilde\rho$-independent $u$, $w$, and $\kappa$ is not possible once the scaling solution for $\kappa(\tilde\rho)$ is imposed.
In general, the flow kernels involve derivatives $\p_{\tilde\rho} u$ etc., such that the rhs of \cref{eq:uExpansionSmall2,eq:wExpansionSmall2,eq:kappaExpansionSmall2} involve $c_1$ and similar terms.
The system is not closed.
The expansion of the scaling equation in powers of $\tilde\rho$ determines the coefficients of terms in higher order as functions of the lower order ones.
This leaves free boundary values or initial values, as $u_0$, $w_0$, and $\kappa_0$, for the local scaling solution in the vicinity of $\tilde\rho$.
For global scaling solutions at least one these UV initial values has to match with the IR initial value $B$ for the local scaling solution for large $\tilde\rho$.

\begin{figure}
\centering
\includegraphics[width=0.50
\columnwidth]{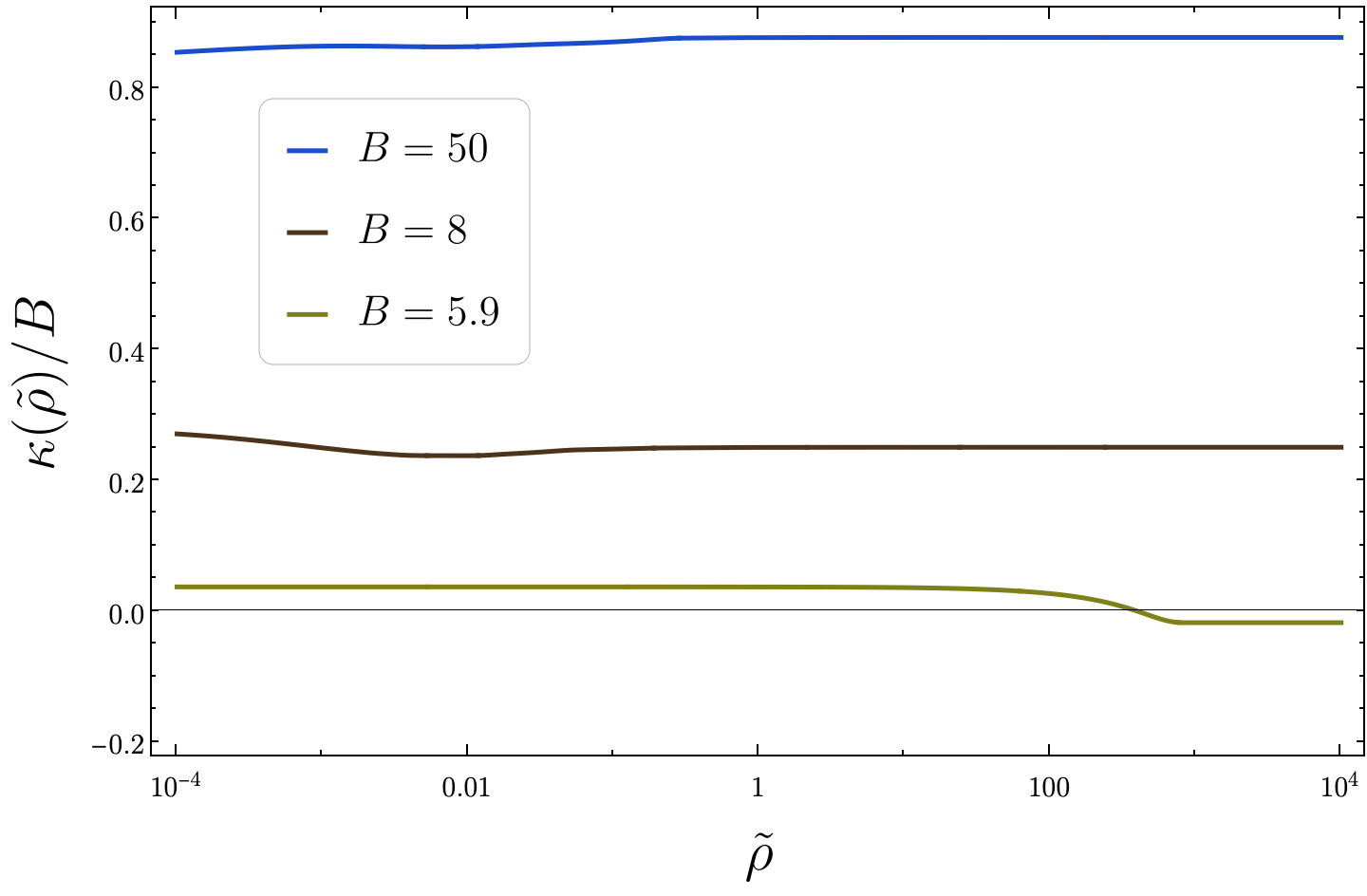}
\caption{
    Scaling solutions for $\kappa(\tilde\rho)/B$ with several values of $B=6+\varepsilon_\infty=6+\kappa_\infty/\xi_\infty$.
    The initial conditions are set at $\tilde\rho_0=10^4$ to be $\kappa(\tilde\rho_0)=\kappa_\infty=\xi_\infty(B-6)$, $u(\tilde\rho_0)=u_\infty =3/(128\pi^2)\approx 0.00238$, $\p_{\tilde\rho}u(\tilde\rho_0)\approx 0$, $w(\tilde\rho_0)= w_0 + \xi_\infty\tilde\rho_0$, and $\p_{\tilde\rho}w(\tilde\rho_0)\approx \xi_\infty$ with $w_0=0.0277$ and $\xi_\infty=1$.
    For all these values of $B$ one finds only small shifts in the values of $\kappa$ for large and small $\tilde\rho$.
}
    \label{fig:u and w and k scaling solution 3}
\end{figure}

One may ask if in our truncation a definite value  of $B$ is singled out, or if scaling solutions exist for continuous ranges in $B$.
For this purpose we solve the scaling equation for different initial values $B$.
A requirement for an acceptable solution is stability of the scalar kinetic term for all values of $\tilde\rho$.
In particular, $\kappa_0=\kappa(\tilde\rho=0)$ has to be positive.
Stability requires positivity of the frame invariant combination $\hat K$,
\al{
\hat K &=\frac{K}{F} + \frac{3}{2F^2}\left(\frac{\p F}{\p \phi}\right)^2\,,&
\hat K k^2&= \frac{1}{2w}\left(\kappa + \frac{6\tilde\rho}{w}\left( \frac{\p w}{\p \tilde\rho}\right)^2 \right)\,.
}
(In the Einstein frame one has $\p F/\p\phi =0$ and $\hat K=K/M_\text{pl}^2$.)
For $\tilde\rho\to 0$ the positivity of $\hat K k^2$ requires $\kappa_0>0$, while for $\tilde\rho\to \infty$ a positive $\hat Kk^2$ is realized for $\kappa_\infty/\xi_\infty + 6>0$ or $B>0$.
In \Cref{fig:u and w and k scaling solution 3} we display $\kappa(\tilde\rho)$ for different values of $B>B_\text{c}$, $B_\text{c}\approx 5.7$ for which $\hat Kk^2$ remains positive for all $\tilde\rho$.
For $B<B_\text{c}$, the stability criterion is violated and numerical instabilities develop.

\begin{figure}
\centering
\includegraphics[width=0.48\columnwidth]{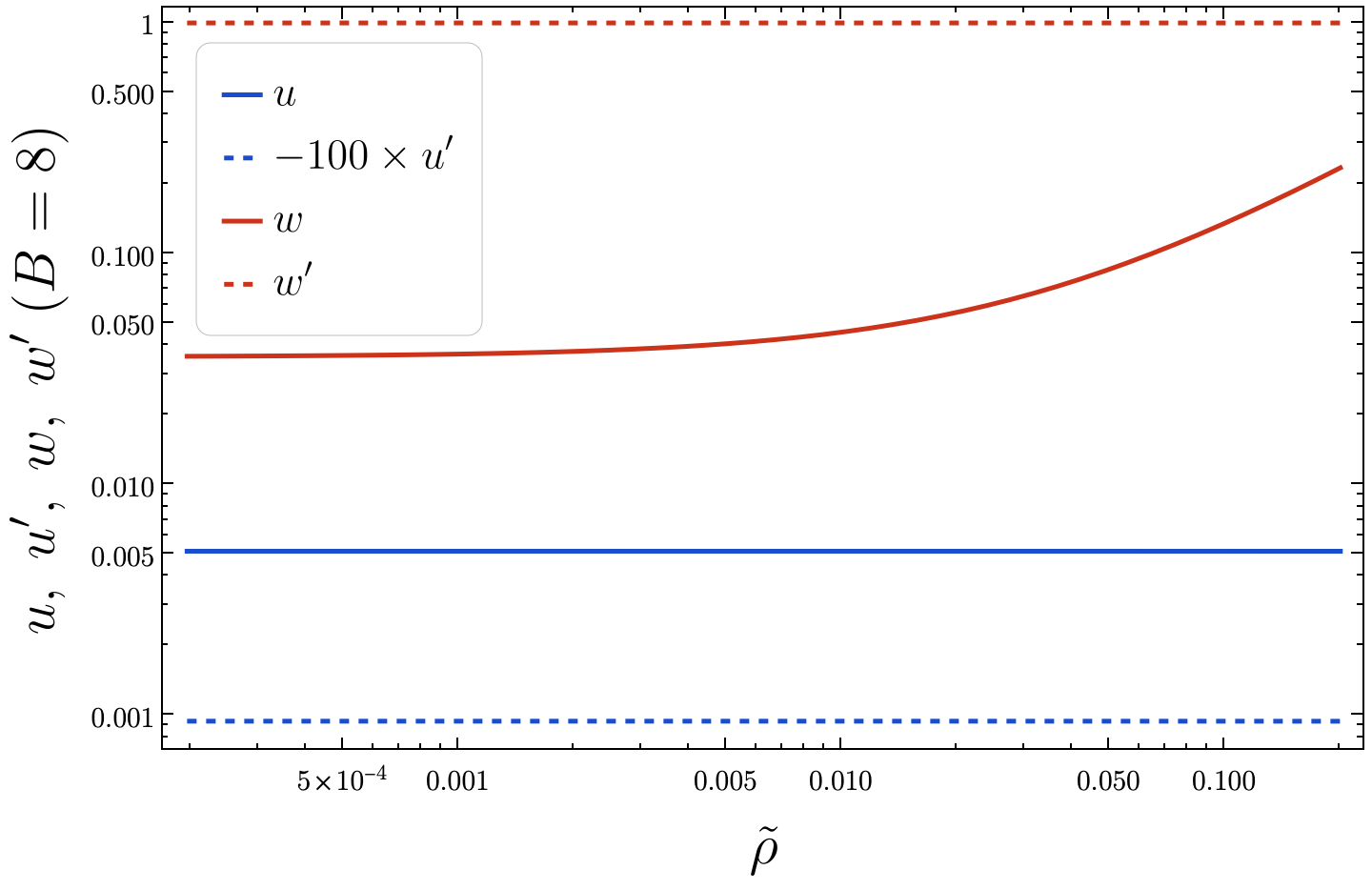}
    \hspace{2ex}
\includegraphics[width=0.48\columnwidth]{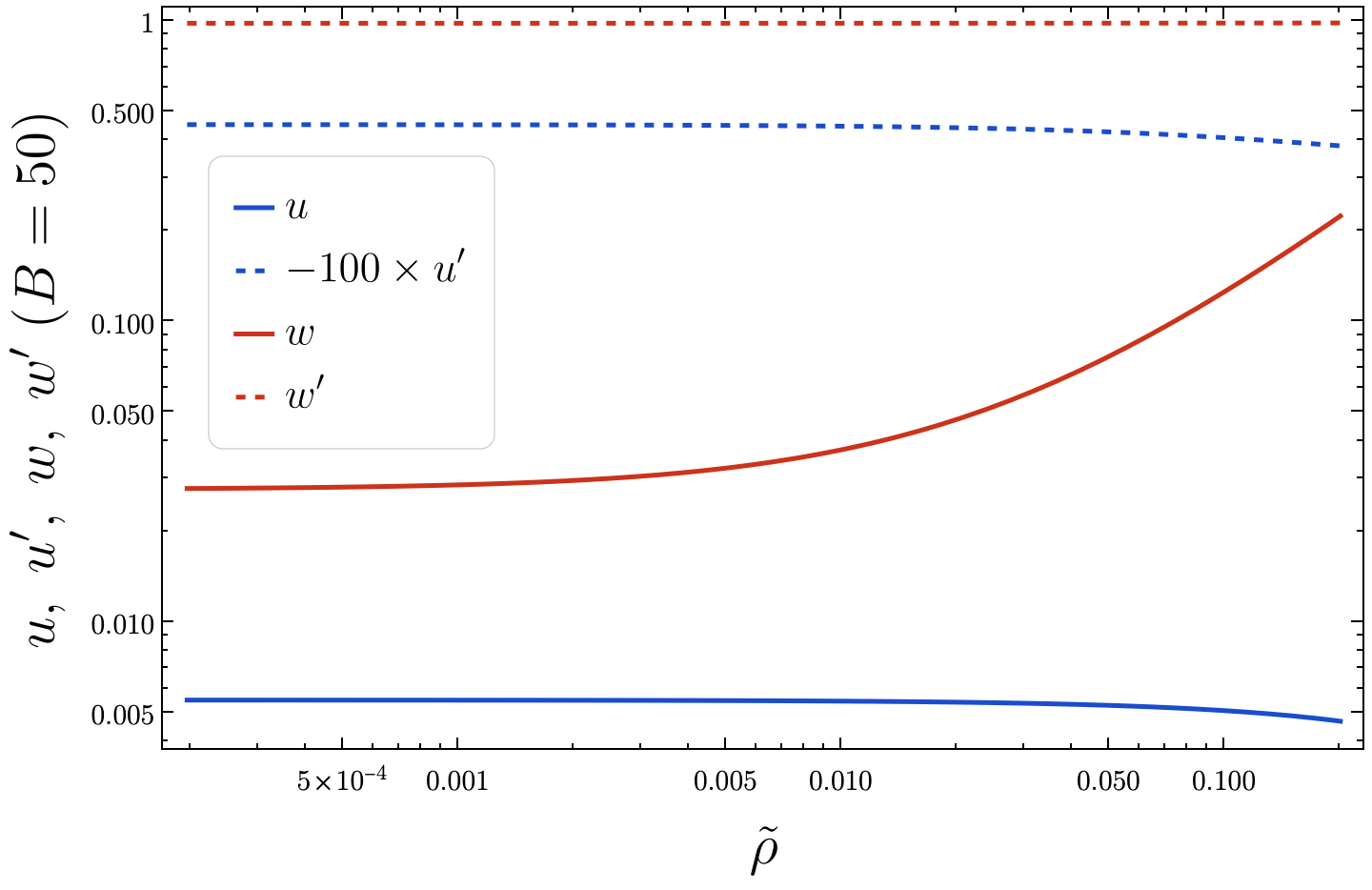}
\caption{
    Scaling solutions for $u(\tilde\rho)$, $u'(\tilde\rho)$, $w(\tilde\rho)$, and $w'(\tilde\rho)$ in cases of $B=8$ (left) and $B=50$ (right).
    The initial conditions are set at $\tilde\rho_0=10^4$ to be $\kappa(\tilde\rho_0)=\kappa_\infty=\xi_\infty(B-6)$, $u(\tilde\rho_0)=u_\infty =3/(128\pi^2)\approx  0.00238$, $\p_{\tilde\rho}u(\tilde\rho_0)\approx 0$, $w(\tilde\rho_0)= w_0 + \xi_\infty\tilde\rho_0$ and $\p_{\tilde\rho}w(\tilde\rho_0)\approx \xi_\infty$ with $w_0=0.0277$ and $\xi_\infty=1$.
    The values of $\tilde\rho\to 0$ differ, most pronounced for $u'(0)$.
}
    \label{fig:u and w and k scaling solution 4}
\end{figure}

Solutions with different $B$ lead to different $u_0$ and $w_0$.
The derivatives $u'(\tilde\rho=0)$ and $w'(\tilde\rho=0)$ are nonzero and depend on $B$.
In \Cref{fig:u and w and k scaling solution 4} we display $u(\tilde\rho)$, $w(\tilde\rho)$, $u'(\tilde\rho)$, and $w'(\tilde\rho)$ for two different values of $B$.
This demonstrates the point made above that for the values $u_0$ and $w_0$ the derivatives of $u$ and $w$ play a role.
At this stage it seems not excluded that within our truncation (and numerical accuracy) continuous families of scaling solutions exist.

\section{Scalar field anomalous dimension}

For the extended Reuter fixed point with $\tilde\rho$-independent $u$ and $w$ one has to change the scaling field by including an anomalous dimension.
In this case $\kappa_0$ becomes a $k$-dependent function.
The relation,
\al{
k\p_k \kappa_0 = -\eta\,\kappa_0\,,
}
defines the anomalous dimension $\eta$.
Switching to renormalized fields $\phi_R=K^{1/2}\phi$, $\tilde\rho_R=K \tilde\rho$ the flow equation \labelcref{eq: beta function of U} at fixed $\tilde\rho_R$ (instead of fixed $\tilde\rho$) reads
\al{
\Big(\p_t - (2+\eta)\tilde\rho\, \p_{\tilde\rho}\Big)u = -4u + \frac{1}{32\pi^2}M_U\,,
}
and similar for the other flow equations.
The form of the scaling equations remains the same if we use
\al{
\p_\tau = -(2+\eta)\tilde\rho \,\p_{\tilde\rho}\,.
}

The scaling equations \labelcref{eq:uExpansionSmall2,eq:wExpansionSmall2} yield the same relations for $u_0$, $w_0$ as before since for $\tilde\rho_R=0$ the lhs $\p_\tau u$ or $\p_\tau w$ vanishes.
For \cref{eq:kappaExpansionSmall2} one replaces $\p_\tau$ by $k\p_k + \p_\tau$ and extracts the anomalous dimension
\al{
\eta = \frac{ (23 w_0-5 u_0)}{240 \pi ^2 (u_0-4 w_0)^2}
 =0.0116\,.
}
This value can be seen as the result of the gauge invariant flow equation for the gravity induced anomalous dimension for scalar fields for the extended Reuter fixed point.
The extended Reuter fixed point is invariant under shifts of the renormalized field $\phi_R\to \phi_R+a$.
This symmetry is preserved in our setting.

For dilaton quantum gravity with the IR fixed point $w=\xi_\infty\tilde\rho$, the modification of the scaling field by an anomalous dimension offers new possibilities for scaling solutions.
The function $\kappa(\tilde\rho)$ may now diverge for $\tilde\rho\to0$, provided that this divergence is absorbed in the definition of a renormalized scaling field $\tilde\rho_R$. 
We will not pursue this possibility in the present paper.
We mention, however, that for strongly diverging $\kappa(\tilde\rho)$ the series expansion for $\tilde\rho\to 0$ may be rearranged.
This is demonstrated by using a different ansatz for $\tilde\rho\to 0$
\al{
\label{eq:IRansatz}
&u(\tilde\rho) = u_0 \,,&
&w(\tilde\rho) = w_0 \,,
&\kappa(\tilde\rho)=\frac{\kappa_{-1}}{\tilde\rho} + \kappa_0 \,.
}
The scaling equations are now given by
\al{
  \label{eq:uExpansionSmall}
&\p_\tau u(\tilde\rho)=
  -4u_0 -\frac{5 u_0}{24 \pi ^2 (u_0-w_0)} -\frac{3 u_0}{160 \pi ^2 (u_0-4 w_0)}  +\frac{1}{96\pi^2} +\frac{79}{480 \pi ^2}\,,
  \\[1ex]
  \label{eq:wExpansionSmall}
&\p_\tau w(\tilde\rho)=
 -2w_0-\frac{25 u_0}{64 \pi ^2 (u_0-w_0)} + \frac{5 u_0^2}{1344 \pi ^2 (u_0-4 w_0)^2} -\frac{9 u_0}{896 \pi ^2 (u_0-4 w_0)}-\frac{1}{192\pi^2}+  \frac{1207}{2688 \pi ^2}\,,
}

\begin{figure}
    \centering
    \includegraphics[width=0.6\columnwidth]{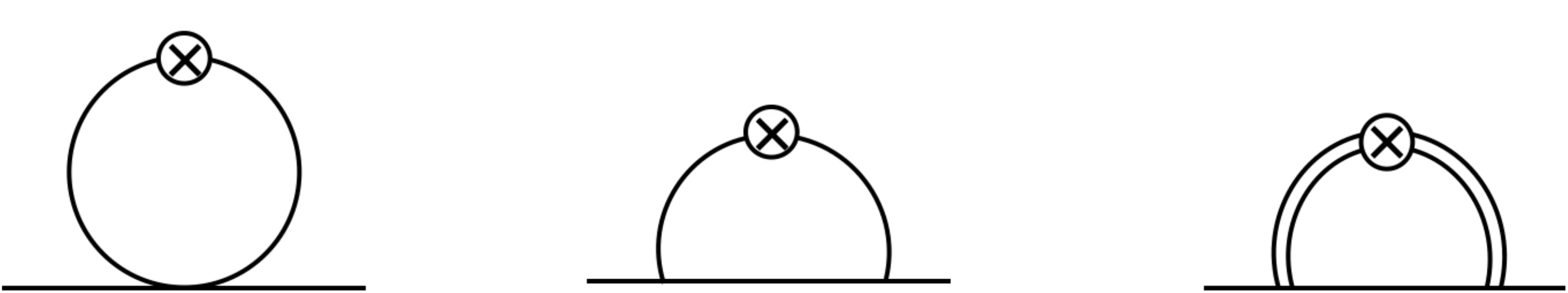}
    \caption{
Diagrams contributing to the flow equation for the kinetial. The single solid line denotes the scalar field $\phi$, while the double solid line is the $\tilde\sigma$ component of the metric. 
The cross in a circle denotes the regulator insertion $\p_t\mathcal R_k$.
The tadpole type diagrams with metric fluctuations [$t_{\mu\nu}$ (traceless-transverse tensor), $\tilde\sigma$ (physical scalar metric fluctuation), $\upsilon_\mu$ (transverse vector), and $u$ (measure scalar)] cancel due to the property \labelcref{eqapp:trace of H2} in \Cref{eq: Quantum corrections to scalar kinetic term}.}
    \label{fig:diag}
\end{figure}

To find the UV-fixed point for $u_0$ and $w_0$ in the $\tilde\rho\to 0$ limit, we solve
\al{
&-4u_0 -\frac{5 u_0}{24 \pi ^2 (u_0-w_0)} -\frac{3 u_0}{160 \pi ^2 (u_0-4 w_0)} + \frac{1}{96\pi^2} +\frac{79}{480 \pi ^2}=0\,,
  \\[1ex]
& -2w_0-\frac{25 u_0}{64 \pi ^2 (u_0-w_0)} - \frac{17 u_0^2-108 u_0 w_0}{2688 \pi ^2 (u_0-4 w_0)^2}- \frac{1}{192\pi^2} +\frac{1207}{2688 \pi ^2}=0\,,
}
where $\p_\tau u_0=\p_\tau w_0=0$ in the left-hand side of \cref{eq:uExpansionSmall,eq:wExpansionSmall}.
We find a UV fixed point
\al{
&u_{0*} = 0.00587\,,&
&w_{0*}= 0.0278\,.
\label{eq: UV fixed point in u and w}
}
For the flow of $\kappa(\tilde\rho)$ the ansanz \labelcref{eq:IRansatz} yields
\al{
\label{eq:kappaExpansionSmall}
&(\p_t+ \p_\tau)\kappa(\tilde\rho) =\frac{1}{(4\pi)^2}\Bigg[
  \left( \frac{9 \kappa_0^2 \tilde{\rho}}{\kappa_{-1}^2} - \frac{25 \kappa_0}{24\kappa_{-1}} + \frac{2}{3 \tilde{\rho}} \right) 
 + \left( -\frac{5 \kappa_0^2 \tilde{\rho}}{2 \kappa_{-1}^2} + \frac{3 \kappa_0}{2 \kappa_{-1}} - \frac{2}{3 \tilde{\rho}} \right) \nn
 &\phantom{(\p_t+ \p_\tau)\kappa(\tilde\rho) =\frac{1}{(4\pi)^2}\Bigg[
  \left( \frac{ \kappa_0^2 \tilde{\rho}}{\kappa_{-1}^2}  \right)}
+\frac{5 u_0 - 23 w_0}{15 (u_0 - 4 w_0)^2}\kappa_0 + \frac{65 u_0 - 296 w_0}{180(u_0 - 4 w_0)^2} \frac{\kappa_{-1} }{\tilde{\rho}}\Bigg]\,.
}
The first two brackets on the rhs correspond to the two graphs in \Cref{fig:diag} with a scalar loop.
The two contributions $2/(3\tilde\rho)$ cancel, consistent with the absence of a term $\sim \tilde\rho^{-1}$ in a pure scalar theory \cite{Wetterich:2024ivi}.
The last term shows a gravitational contribution $\sim \tilde\rho^{-1}$.
This leads to an anomalous dimension slightly different from two.

From \cref{eq:kappaExpansionSmall} we extract the flow of $\kappa_{-1}$ from the term $\sim \tilde\rho^{-1}$
\al{
(\p_t + 2) \kappa_{-1} = -\frac{1}{(4\pi)^2}\frac{(296 w_0-65 u_0 )}{180(u_0 - 4 w_0)^2} \kappa_{-1}\,,
}
which yields
\al{
\tilde\eta =- \left( 2 +  \frac{1}{(4\pi)^2} \frac{(296 w_0-65 u_0 )}{180(u_0 - 4 w_0)^2} \right) 
\approx -2.026\,.
}

For dilaton quantum gravity with $w\sim \xi\tilde\rho$ for large $\tilde\rho$ a nonzero anomalous dimension for $\tilde\rho\to0$ would have to be combined with a zero anomalous dimension for $\tilde\rho \to \infty$.
Thus the anomalous dimension should depend on $\tilde\rho$.
This requires a nonlinear relation between the renormalized scalar field $\phi_R$ and $\phi$.
Nonlinear scaling fields are possible.
Nonlinear field transformations induce, however, additional terms in the flow equation \cite{Wetterich:1996kf,Gies:2001nw,Gies:2002hq,Pawlowski:2005xe,Floerchinger:2009uf,Baldazzi:2021ydj,Knorr:2022ilz,Ihssen:2022xjv,Wetterich:2024uub,Falls:2025sxu}.

\section{Conclusions}

We have investigated the scaling solution for dilaton quantum gravity in the variable gravity approximation.
It defines an ultraviolet complete quantum field theory for a scalar singlet coupled to the metric, as used for models of inflation or dynamical dark energy.
We employ the gauge invariant flow equation, which focuses on physical fluctuations, respecting diffeomorphism symmetry and global dilatation symmetry.

Our results for the overall features of the dimensionless scalar potential $u$ and the dimensionless squared Planck mass $w$ are in accordance with earlier investigations.
As a function of the squared dimensionless scalar field $\tilde\rho$ one finds an almost flat potential $u(\tilde\rho)$ and a Planck mass increasing for large values of the scalar field, $w(\tilde\rho)\approx w_0 + \xi_\infty \tilde\rho$.
These central properties of the dilaton quantum gravity fixed point do not depend on the particular version of the functional flow equation or on the precise truncation. 
This enhances the robustness of the arguments in favor of the existence of such an ultraviolet fixed point.

The use of the gauge invariant flow equation may contribute to a better understanding of the field-dependence of the kinetial---the coefficient in front of the scalar kinetic term.
Because of its focus on the physical fluctuations it can take well into account the mixing of the kinetic terms for the scalar field and the physical scalar fluctuation in the metric.
In particular, it allows us to explore a negative sign of the kinetial in the region of large $\tilde\rho$, as long as it remains consistent with a stability criterion requiring a positive kinetic term for the eigenstate corresponding to the inflaton or cosmon.

Within our approximation we find two different types of scaling solutions, corresponding to two different ultraviolet fixed points.
For the extended Reuter fixed point the potential $u(\tilde\rho)$ and the squared Planck mass $w(\tilde\rho)$ are independent of $\tilde\rho$.
Such a scaling solution employs a renormalized scaling field.
The dependence of $K$ on $k$ involves a nonvanishing anomalous field dimension $\eta$.
We compute the quantum gravity contribution to $\eta$.

The second UV-fixed point is the dilaton quantum gravity fixed point with $w(\tilde\rho)$ increasing $\sim \tilde\rho$ for large $\tilde\rho$.
Our truncation and numerical solution seem to be compatible with a continuous family of scaling solutions parametrized by $B$.
This family includes the possibility of a negative kinetial for large $\tilde\rho$.
For the scaling solutions found in the present paper the anomalous field dimension $\eta$ vanishes for all $\tilde\rho$.
It remains to be seen if further scaling solutions can be found with $\eta$ depending on $\tilde\rho$, vanishing only for $\tilde\rho\to \infty$.
Such scaling solutions require a scaling field which is related to the original or microscopic scalar field by a nonlinear field transformation.
On the other hand, more extended truncations may reduce the continuous family of scaling solutions to a discrete set, leave only a single scaling solution, or even exclude the existence of scaling solutions with $w(\tilde\rho)\sim \tilde\rho$ for large $\tilde\rho$.

At the present stage the argument for the existence of a dilaton quantum gravity fixed point different from the extended Reuter fixed point is strengthened by our investigation.
This fixed point realizes the exact quantum scale symmetry in the infrared limit for $\tilde\rho\to\infty$.
The issue of quantum scale symmetry for the ultraviolet limit $\tilde\rho\to0$ will depend on the anomalous field dimension.

\begin{acknowledgments}
Y.\,M and M.\,Y. thank the institute for Theoretical Physics, Heidelberg University and Tokyo Women's Christian University for kind hospitality during their stay.
The work of M.\,Y. was supported by the National Science Foundation of China (NSFC) under Grant No.\,12205116 and the Seeds Funding of Jilin University. 
\end{acknowledgments}

\appendix
\section{Effective average action for nonminimally coupled scalar-gravity system}
\label{formulations}

The central method in this work is the functional renormalization group (FRG)~\cite{Wilson:1973jj}. For quantum field theory we employ a functional differential equation for the effective average action~\cite{Wetterich:1992yh} (see also \cite{Morris:1993qb,Ellwanger:1993mw,Reuter:1993kw})
\al{
  \p_t \Gamma_k[\Phi] = \frac{1}{2}\Tr \left[
    \left({\Gamma^{(2)}_k [\Phi] + {\mathcal R}_k } \right)^{-1}\, \p_t {\mathcal R}_k\right]\,.
  \label{eq:wetterich}
}
Here $\p_t=k\p_k$ is a dimensionless scale derivative and ``STr" in \cref{eq:wetterich} is summing over all internal degrees of freedom and momenta of the superfield $\Phi$. 
The matrix of second functional derivatives $\Gamma^{(2)}_k$ corresponds to the full two-point function, i.e., full inverse propagator of $\Phi$ in which an infrared cutoff profile function ${\mathcal R}_k$ is introduced.

\subsection{Setup}

In this work, we set up the system of a singlet scalar field $\phi$ nonminimally coupled to gravity using the FRG equation \labelcref{eq:wetterich}. In such a system, we give the effective action as
\al{
  \Gamma_k&=\Gamma_k^{\rm SG} +\Gamma_{\rm gf} +\Gamma_{\rm gh}\,.
  \label{effective action for both}
}
Here, our truncated scalar-gravity part $\Gamma_k^{\rm SG}$ is given by
\al{
&\Gamma_k^{\rm (SG)}
	=	\int_x \sqrt{g}\left\{U(\rho)
		+ \frac{K(\rho)}{2} g^{\mu\nu} \p_\mu{\phi}\p_\nu\phi -\frac{F(\rho)}{2}R 
		\right\}\,.
				\label{effective action for scalar}
}
We have defined the field dependent Planck mass $F(\rho)$, the scalar effective potential $U(\rho)$ and the kinetic potential $K(\rho)$ as functions of the scalar invariant $\rho=\phi^2/2$.

To specify the physical gauge fixing for diffeomorphism acting on the metric fluctuation field $h_{\mu\nu}$, we expand the macroscopic metric  around a constant background metric  ${\bar g}_{\mu\nu}$ such that
\al{
g_{\mu\nu}={\bar g}_{\mu\nu}+h_{\mu\nu}\,. 
}
In this work, we employ the maximally symmetric background space for simplicity, i.e.,
\al{
&\bar R_{\mu\rho\nu\sigma} = \frac{\bar R}{12} \Big( \bar g_{\mu\nu} \bar g_{\rho\sigma} - \bar g_{\mu\sigma} \bar g_{\nu\rho}\Big)\,,&
&\bar R_{\mu\nu}=\frac{\bar R}{4} \bar g_{\mu\nu}\,.
\label{eq: maximally symmetric space}
}
In the current setup for the scalar-gravity sector~\labelcref{effective action for scalar}, this choice provides the same result of the beta functions as the case with a general homogeneous background field. The gauge fixing for diffeomorphism is given by
\al{
 \Gamma_{\rm gf} &=
                        \frac{1}{2\alpha}\int_x\sqrt{\bar g}\,
                        {\bar g}^{\mu \nu}\Sigma_\mu\Sigma_{\nu} \,.
                        \label{gauge fixing action}
}
Here, the general gauge fixing reads
\al{
\Sigma_\mu		
	= {\bar D}^\nu h_{\nu \mu}-\frac{\beta +1}{4}{\bar D}_\mu h\,,
	\label{gauge fixing function}
}
with $h={\bar g}^{\mu\nu}h_{\mu\nu}$ the trace mode of metric fluctuations.
Bars on operators denote that covariant derivatives and index contractions are formed with the background metric.
The ghost action associated with the gauge fixing \labelcref{gauge fixing function} is given by 
\al{
  \Gamma_{\rm gh}	&=	-\int_x\sqrt{\bar g}\,\bar C_\mu
               \left[ {\bar g}^{\mu\rho}{\bar D}^2+
               \frac{1-\beta}{2}{\bar D}^\mu{\bar D}^{\rho}
               +{\bar R}^{\mu\rho}\right] C_{\rho}\,, \label{ghostaction}
}
where $C$ and $\bar C$ are ghost and antighost fields. 
\Cref{gauge fixing action,gauge fixing function} constitute a general family of gauge fixings for diffeomorphism symmetry, specified by two parameters $\alpha$ and $\beta$.
The choice $\beta=-1$ and $\alpha\to 0$ is a ``physical gauge fixing'' that acts only on the gauge modes in $h_{\mu\nu}$, leaving the physical fluctuations unaffected.
In this work, we use this gauge choice.
Nevertheless, the following subsection treats general $\alpha$ and $\beta$ in order to demonstrate explicitly the particular role of the physical gauge fixing.

\subsection{Physical metric decomposition}

A key quantity for the flow equation is the inverse propagator or one-particle-irreducible two-point function, as given by the matrix of second functional derivatives of $\Gamma_k$. 
To derive the explicit form for the metric two-point function, we split the metric fluctuations into physical and gauge fluctuations~\cite{Wetterich:2016vxu}. 
Accordingly, we decompose the metric fluctuations into
\al{
h_{\mu\nu}=f_{\mu\nu}+a_{\mu\nu}\,,
\label{metric decomposition}
}
where $f_{\mu\nu}$ are the physical metric fluctuations, while $a_{\mu\nu}$ are the gauge fluctuations or gauge modes.
The physical metric fluctuations satisfy the transverse constraint $\bar D^\mu f_{\mu\nu}=0$. 
In turn, the physical metric fluctuations can be decomposed into two independent fields as
\al{
f_{\mu\nu}=t_{\mu\nu}+s_{\mu\nu}\,,
\label{physical decomposition}
}
where the graviton $t_{\mu\nu}$ is the transverse and traceless (TT) tensor, i.e., $\bar D^\mu t_{\mu\nu}=\bar g^{\mu\nu}t_{\mu\nu}=0$.
The tensor $s_{\mu\nu}$ is given as a linear function of a scalar field $\sigma$ such that
\al{
s_{\mu\nu}=\hat S_{\mu\nu} \sigma =\frac{1}{3}P_{\mu\nu} \sigma\,.
}
For a background geometry with constant curvature, the projection operator can be found explicitly as
\al{
&P_{\mu\nu} =\bigg( \bar g_{\mu\nu}\bar \Delta_S + \bar D_\mu \bar D_\nu - \bar R_{\mu\nu}\bigg) \left(\bar \Delta_S-\frac{\bar R}{3}\right)^{-1}
= \bigg[ \bar g_{\mu\nu}\left( \bar \Delta_S - \frac{\bar R}{4}  \right) + \bar D_\mu \bar D_\nu \bigg] \left(\bar \Delta_S-\frac{\bar R}{3}\right)^{-1}\,,
\label{projection operator Pmunu}
}
with $\bar \Delta_S=-\bar D^2=-\bar D^\mu \bar D_\mu$ the covariant Laplacian acting on scalar fields (spin-0 fields). Here,  in the second equality, we have used the maximally symmetric space \labelcref{eq: maximally symmetric space}.

Similarly, the gauge modes or unphysical metric fluctuations $a_{\mu\nu}$ are decomposed into a transverse vector mode $\upsilon_\mu$, satisfying $\bar D^\mu \upsilon_\mu=0$, and a scalar mode $u$. In summary, the metric fluctuations \labelcref{metric decomposition} are parametrized by
\al{
f_{\mu\nu}&=t_{\mu\nu}+\frac{1}{3}P_{\mu\nu} \sigma\,, &
a_{\mu\nu}&=\bar D_\mu \upsilon_\nu+\bar D_\nu \upsilon_\mu - \bar D_\mu  \bar D_\nu  u\,.
\label{parametrized metric}
}
Note here that the trace mode of the metric fluctuations $h=\bar g^{\mu\nu}h_{\mu\nu}$ is given by
\al{
h = \bar g^{\mu\nu} \left(t_{\mu\nu} +\frac{1}{3}P_{\mu\nu} \sigma  + \bar D_\mu \upsilon_\nu+\bar D_\nu \upsilon_\mu - \bar D_\mu  \bar D_\nu  u\right)
= \sigma + \bar\Delta_S u\,.
}

The ghost fields can be decomposed similarly into vector and scalar fields
\al{
C_\mu&=C_{\mu}^\perp+\bar D_\mu C\,,&
\bar C_\mu&=\bar C_{\mu}^\perp+\bar D_\mu \bar C\,,
\label{parametrized ghosts}
}
where $C_{\mu}^\perp$ ($\bar C_{\mu}^\perp$) is the transverse (anti)ghost field and $C$ ($\bar C$) is the scalar (anti)ghost field.

\subsection{Jacobians}

The decompositions of the metric fluctuation \labelcref{parametrized metric} and the ghost fields \labelcref{parametrized ghosts} yield Jacobians of the path integral measures for different modes.
We start with the Gaussian integral for the metric fluctuations $h_{\mu\nu}$. In the maximally symmetric space \labelcref{eq: maximally symmetric space}, the Gaussian integrals for each mode in the physical metric decomposition read
\al{
1&=\int {\mathcal D}h_{\mu\nu}\,\exp\fn{-\frac{1}{2}h_{\mu\nu}h^{\mu\nu}}\nn
&=J_\text{grav}\int {\mathcal D}t_{\mu\nu} {\mathcal D}\sigma {\mathcal D \upsilon_\mu} \mathcal D u \,\exp\bigg[ -\frac{1}{2}\int_x \sqrt{\bar g}\,
\bigg\{
t_{\mu\nu} t^{\mu\nu}
+ \sigma \hat S^{\alpha\beta}\hat S_{\alpha\beta} \sigma
+2\upsilon_\mu \left( \bar{\mathcal D}_\text{1} \right)^{\mu\nu} \upsilon_\nu
+2u \bar{\mathcal D}_\text{0} \bar\Delta_S u
\bigg\}\bigg]\nn[1ex]
&=J_\text{grav}\left[ \det{}'_\text{(1T)}\left(\bar{\mathcal D}_\text{1} \right) \right]^{-1/2}
\left[ \det{}'_{(0)}\left( \bar{\mathcal D}_0\right)\bar \Delta_S  \right]^{-1/2}
\left[ \det{}'_{(0)}\left(\hat S^{\alpha\beta}\hat S_{\alpha\beta} \right)\right]^{-1/2}\,,
}
where $E_{(t)}$ is the identical matrix acting on the space satisfying the TT condition.
We have in the maximal symmetric space and $\bar g^{\mu\nu}t_{\mu\nu}=0$.
One finds the Jacobian arising from the decomposition of the metric fluctuation,
\al{
J_\text{grav}&=\left[ \det{}_\text{(1T)}'\left(\bar{\mathcal D}_\text{1} \right) \right]^{1/2}
\left[ \det{}_{(0)}'\left(\bar{\mathcal D}_0\right)\bar \Delta_S\right]^{1/2}
\left[ \det{}'_{(0)}\left(\hat S^{\alpha\beta}\hat S_{\alpha\beta} \right)\right]^{1/2}
\equiv J_{\upsilon} \cdot J_{u} \cdot J_{\sigma}
\,.
\label{App: Jacobian from metric decomposition}
}
with differential operators
\al{
\bar{\mathcal D}_1&=\bar \Delta_V-\frac{\bar R}{4}\,,&
\bar{\mathcal D}_0&=\bar \Delta_S-\frac{\bar R}{4}\,.
\label{appeq: D1 and D0}
}
Here $\bar \Delta_V=-\bar D^2$ is the Laplacian acting on vector fields (spin-1 fields).
A prime denotes a subtraction of a zero eigenmode. This subtraction, however, does not contribute to the beta functions within the present truncation, so that we hereafter neglect it.
The Jacobian for $\sigma$ in the maximally symmetric space \labelcref{eq: maximally symmetric space} can be computed as
\al{
J_{\sigma}&=  \left[\det{}_\text{(0)} \left( \hat S^{\alpha\beta}\hat S_{\alpha\beta} \right)\right]^{1/2}
= \left[\det{}_\text{(0)} \frac{1}{3}\left( \bar\Delta_S -\frac{\bar R}{3} \right)^{-1}  \bar{\mathcal D}_0 \right]^{1/2}\,.
\label{eq: Jacobian for sigma}
}
For the evaluation of fluctuations of $\sigma$, it is convenient to remove $\left( \bar\Delta_S -\frac{\bar R}{3} \right)^{-1}$. This can be done by redefining the field $\sigma$ such that
\al{
\tilde\sigma = \left( \bar\Delta_S -\frac{\bar R}{3}\right)^{-1}\sigma\,,
\label{app:eq: redefinition of sigma}
}
for which the Jacobian \labelcref{eq: Jacobian for sigma} reads
\al{
J_{\tilde\sigma}= \left[\det{}_\text{(0)}\left( \bar{\mathcal D}_0 \right)\left( \bar\Delta_S -\frac{\bar R}{3}\right) \right]^{1/2}\,.
}

In the same manner, the Gauss integral for the ghost fields is
\al{
1&= \int \mathcal D C_\mu\mathcal D \bar C_\mu \,\exp\left( \bar C_\mu \bar C^\mu\right)
= J_{\rm gh}\int \mathcal D C_\mu^\perp\mathcal D \bar C_\mu^\perp \mathcal D C\mathcal D \bar C \,\exp\left( \bar C^\perp_\mu \bar C^{\perp\mu} +\bar C \bar\Delta_S C\right)
=J_{\rm gh} \left[\det{}_{(0)} (\bar\Delta_S) \right]\,,
}
namely one has
\al{
J_{\rm gh}= \left[ \det{}_{(0)}\left(\bar \Delta_S \right) \right]^{-1}\,.
}

To summarize, the decompositions yield Jacobians are found to be
\al{
J_{\upsilon}&=\left[ \det{}_\text{(1T)}'\left(\bar{\mathcal D}_1 \right) \right]^{1/2}\,, &
J_{\tilde\sigma}&= \left[ \det{}_{(0)}'\left(\bar{\mathcal D}_0\right)\left( \bar\Delta_S -\frac{\bar R}{3}\right)\right]^{1/2}\,,\nn
J_{u}&= \left[ \det{}_{(0)}'\left(\bar{\mathcal D}_0\right)\bar \Delta_S\right]^{1/2}\,,&
J_\text{gh}&=\left[ \det{}_{(0)}''\left(\bar \Delta_S \right) \right]^{-1}\,.
 \label{set of Jacobians}
}
By the same reason in \cref{App: Jacobian from metric decomposition}, we neglect the primes denoting the subtraction of zero modes.
For the regularized Jacobians we replace the relevant differential operator $\bar{\mathcal D}$ by $P_k(\bar{\mathcal D})$ to avoid the Jacobians to induce strong nonlocalities. 
Then, these Jacobians are written in forms of flow equations as
\al{
&J_{\upsilon} = -\frac{1}{2}\Tr_{\rm (1T)}\frac{\p_t \mathcal R_k(\bar{\mathcal D}_1)}{P_k(\bar{\mathcal D}_1 )}\,,
\label{app: eq: kappa Jacobian}
\\
&J_{\tilde\sigma} = -\frac{1}{2}\Tr_{(0)}\frac{\p_t \mathcal R_k(\bar{\mathcal D}_0)}{P_k(\bar{\mathcal D}_0 )}
-\frac{1}{2}\Tr_{(0)}\frac{\p_t \mathcal R_k(\bar\Delta_S -\frac{\bar R}{3})}{P_k(\bar\Delta_S -\frac{\bar R}{3} )}\,,
\label{app eq: Jacobian for sigma tilde}
\\
&J_{u}=  -\frac{1}{2}\Tr_{(0)}\frac{\p_t \mathcal R_k(\bar{\mathcal D}_0)}{P_k(\bar{\mathcal D}_0 )}  -\frac{1}{2}\Tr_{(0)}\frac{\p_t \mathcal R_k(\bar \Delta_S)}{P_k(\bar \Delta_S )} \,,
\label{app eq: Jacobian for u}
\\
&J_\text{gh}= \Tr_{(0)}\frac{\p_t \mathcal R_k(\bar \Delta_S )}{P_k(\bar \Delta_S )}\,.
\label{app eq: Jacobian for ghost}
}
%

\section{Variations}
\label{app: variations}

To obtain the Hessian, we need to compute the variations for the effective action~\labelcref{effective action for both}. Hence, we expand the effective action around background fields into polynomials of fluctuation fields:
\al{
\Gamma_k[\Phi] = \Gamma_k[\bar\Phi] + \Gamma_k^{(1)}[\bar\Phi](x)\cdot  \delta\Phi(x) +\frac{1}{2}\delta\Phi(x)\cdot \Gamma^{(2)}_k[\bar\Phi](x,y)\cdot  \delta\Phi(y) + \cdots\,,
}
from which the Hessian is identified as $\Gamma^{(2)}_k[\bar\Phi](x,y) = \frac{\delta^2\Gamma_k}{\delta \Phi(x) \delta \Phi(y)}|_{\Phi=\bar\Phi}$.
Here, the metric field and the scalar field are linearly expanded around their respective background fields $\bar g_{\mu\nu}$ and $\bar\phi$:
\al{
&g_{\mu\nu} = \bar g_{\mu\nu} + h_{\mu\nu}\,,&
&\phi = \bar\phi + \varphi\,.
}
For our effective action
\al{
&\Gamma_k^{\rm (SG)}[g_{\mu\nu},\phi]
=\int_x \sqrt{g}g^{\mu\nu}\frac{K(\rho)}{2} \p_\mu \phi \p_\nu \phi
+ \int_x\sqrt{g}U\fn{\rho}
- \int_x \frac{F\fn{\rho}}{2}\sqrt{g}R\,,
}
the second variation reads
\al{
\delta^2\Gamma_k^{\rm (SG)}[\bar\Phi]
&= \int_x\sqrt{\bar g} \left\{  2\bar g^{\mu\nu}K(\bar\rho)\right\} \frac{1}{2}\p_\mu\varphi \p_\nu \varphi 
+ \int_x\sqrt{\bar g} \left[\{\bar g^{\mu\nu} (\delta^2K(\rho))\big|_{\phi=\bar\phi}\} \frac{1}{2}\p_\mu \bar\phi \p_\nu\bar\phi
+ \{ 4\bar g^{\mu\nu}\delta K(\rho)\big|_{\phi=\bar\phi} \} \frac{1}{2}\p_\mu\varphi \p_\nu \bar\phi
\right]
\nn
&\quad
+ \int_x\sqrt{\bar g} H_2^{\mu\nu} K(\bar\rho)\frac{1}{2}\p_\mu \bar\phi \p_\nu \bar\phi 
+\int_x\sqrt{\bar g} 2H_1^{\mu\nu}\left( 2K(\bar\rho)\frac{1}{2}\p_\mu \varphi \p_\nu \bar\phi + (\delta K(\rho)\big|_{\phi=\bar\phi})\frac{1}{2} \p_\mu \bar\phi \p_\nu\bar\phi  
\right)
\nn
&\quad
+ \int_x \sqrt{\bar g}\left[  G_2 U(\bar\rho) +  2G_1\delta U(\rho) + \delta^2U\fn{\rho}\right]\bigg|_{\phi=\bar\phi}
-  \int_x \sqrt{\bar g}\frac{1}{2} \left[ I_2 F(\bar\rho) + 2 I_1\delta F(\rho)  + \delta^2F\fn{\rho} \bar R \right]\bigg|_{\phi=\bar\phi}\,.
\label{app:eq: second variation}
}
Here, we have defined 
\al{
&H_1^{\mu\nu} =\delta (\sqrt{g}g^{\mu\nu})/\sqrt{\bar g}=\frac{1}{2} h\bar g^{\mu\nu} - h^{\mu\nu}\,,\\[1ex]
&H_2^{\mu\nu} = \delta^2 (\sqrt{g}g^{\mu\nu})/\sqrt{\bar g} = \left( -\frac{1}{2}h^{\alpha\beta}h_{\alpha\beta} +\frac{1}{4} h^2\right)\bar g^{\mu\nu} + 2h^\mu{}_\lambda h^{\nu\lambda}  - hh^{\mu\nu}\,,\\[1ex]
&G_1= \delta\sqrt{g}/\sqrt{\bar g} =\frac{1}{2}h \,,\\[1ex]
&G_2 = \delta^2 \sqrt{g}/\sqrt{\bar g} =-\frac{1}{2} h^{\alpha\beta}h_{\alpha\beta} + \frac{1}{4}h^2 +\bar D_\mu \bar D_\nu h^{\mu\nu} + \bar\Delta_S h\,,  \\[1ex]
&I_1= \delta(\sqrt{g} R)/\sqrt{\bar g}  
=\frac{1}{2} \bar R h - \bar R^{\alpha\beta} h_{\alpha\beta}\,,\\[1ex]
&I_2= \delta^2(\sqrt{g} R)/\sqrt{\bar g} 
= -\frac{1}{2} h^{\alpha\beta} \bar\Delta_T h_{\alpha\beta}  -\frac{1}{2}\bar R h^{\alpha\beta}h_{\alpha\beta}   + h^{\alpha\beta} \bar R_\alpha{}^\rho h_{\rho \beta} + h \bar D^\mu \bar D^\nu h_{\mu\nu} + \bar D^\mu h_{\mu\alpha} \bar D^\nu h_{\nu}{}^\alpha \nn
&\qquad\qquad\qquad\qquad\qquad
+ \frac{1}{2} h\bar\Delta_S h + h^{\alpha\beta}\bar R_\alpha{}^\mu{}_\beta{}^\nu h_{\mu\nu} - h\bar R^{\alpha \beta}h_{\alpha\beta} + \frac{1}{4} \bar R h^2 \,,
}
and the variations for the scalar potentials $K$, $F$ and $U$ are computed as
\al{
&\delta K(\rho)\Big|_{\phi=\bar\phi} =  (K' (\bar\rho)\bar\phi )\varphi\,,&
&\delta^2 K(\rho)\Big|_{\phi=\bar\phi} =(K'(\bar\rho)+2\bar\rho K'' (\bar\rho))\varphi^2\,,\\
&\delta U(\rho)\Big|_{\phi=\bar\phi} = ( U' (\bar\rho)\bar\phi)\varphi \,,&
&\delta^2 U(\rho)\Big|_{\phi=\bar\phi} =(U'(\bar\rho)+2\bar\rho U''(\bar\rho))\varphi^2 \,,\\
&\delta F(\rho) \Big|_{\phi=\bar\phi}=  (F' (\bar\rho)\bar\phi)\varphi\,,&
&\delta^2 F(\rho)\Big|_{\phi=\bar\phi}  =(F'(\bar\rho)+2\bar\rho F''(\bar\rho))\varphi^2\,,
}
where primes denote derivatives with respect to $\bar\rho=\bar\phi^2/2$.

\section{Hessians}

In this section, we focus on the evaluation of the quantum corrections to $F(\bar\rho)$ and $U(\bar\rho)$. We assume a constant background scalar field $\bar\phi$ and therefore drop the term $\p_\mu \bar\phi$ in the Hessians. The quantum corrections to $K(\bar\rho)$ are discussed in \Cref{eq: Quantum corrections to scalar kinetic term}. In the following, to make computations simple, we employ the maximal symmetric space for the background metric \labelcref{eq: maximally symmetric space}. We present the explicit Hessians for each mode:
\al{
\delta^2\Gamma_k^{\rm (SG)}[\bar\Phi]
&= \int_x \sqrt{\bar g}\Big[  \bar g^{\mu\nu}K(\bar\rho)\varphi (- \p_\mu\p_\nu) \varphi \Big]\bigg|_{\phi=\bar\phi}
+ \int_x \sqrt{\bar g}\left[  G_2 U(\bar\rho) +  2G_1\delta U(\rho) + \delta^2U\fn{\rho}\right]\bigg|_{\phi=\bar\phi}\nn
&\quad
- \frac{1}{2} \int_x \sqrt{\bar g}\left[ I_2 F(\bar\rho) + 2 I_1\delta F(\rho)  + \delta^2F\fn{\rho} \bar R \right]\bigg|_{\phi=\bar\phi}\,.
}
Then, the two-point functions (Hessians) $\Gamma_k^{(2)}$ is given by
\al{
\delta^2\Gamma_k[\bar\Phi]
= \delta\Phi \cdot\Gamma_k^{(2)}[\bar\Phi]\cdot \delta\Phi
&= \int_x\sqrt{\bar g}\pmat{h_{\mu\nu} & \varphi}
\pmat{
\left(\Gamma_{(hh)}^{(2)}\right)^{\mu\nu\rho\sigma} & \left(\Gamma_{(h\varphi)}^{(2)}\right)^{\mu\nu} \\[3ex] 
\left(\Gamma_{(\varphi h)}^{(2)}\right)^{\rho\sigma} & \Gamma_{(\varphi\varphi)}^{(2)}
}
\pmat{h_{\rho\sigma}\\[4ex] \varphi}\nn[5ex]
&= \int_x\sqrt{\bar g} \pmat{t_{\mu\nu}  & \upsilon_\mu & \sigma & u & \varphi}
\pmat{
\left(\Gamma_{(tt)}^{(2)}\right)^{\mu\nu\rho\sigma} & 0 & 0 & 0  & 0\\[1ex]
0 & \left(\Gamma_{(\upsilon\upsilon)}^{(2)}\right)^{\mu\nu}  & 0 & 0  & 0\\[1ex]
0 & 0 &  \Gamma_{(\sigma\sigma)}^{(2)}   &  \Gamma_{(\sigma u)}^{(2)} & \Gamma_{(\sigma\varphi)}^{(2)}\\[1ex]
0 & 0 &  \Gamma_{(u\sigma)}^{(2)}   &  \Gamma_{(u u)}^{(2)} & \Gamma_{(u \varphi)}^{(2)}\\[1ex]
0 & 0 &  \Gamma_{(\varphi\sigma)}^{(2)}   &  \Gamma_{(\varphi u)}^{(2)} & \Gamma_{(\varphi\varphi)}^{(2)}
}
\pmat{t_{\rho\sigma}\\[2ex]  \upsilon_\nu \\[2ex] \sigma \\[2ex] u \\[2ex] \varphi}\,.
}
For our decomposition of the metric fluctuations, the matrix of the Hessians becomes block diagonal for each degree of freedom or spin. Note that the Hessians depend on only the background fields $\bar g_{\mu\nu}$ and $\bar\phi$.

\subsection{Transverse-traceless tensor}

For the $t_{\mu\nu}$ mode, we obtain the Hessian as
\al{
\left(\Gamma_{(tt)}^{(2)}\right)^{\mu\nu\rho\sigma}= \left[ \frac{F}{2}\bar{\mathcal D}_T -  U \right]P^{(t)\mu\nu\rho\sigma}\,,
\label{TT mode two point function}
}
where we define the derivative operator
\al{
\bar{\mathcal D}_T=\bar \Delta_T+\frac{2\bar R}{3}\,,
}
with the Laplacian $\bar\Delta_T=- \bar D^2$ acting on transverse traceless tensor fields (spin-2 fields).
The TT-projection operator is given by
\al{
P^{(t)\mu\nu\rho\sigma}=
\frac{1}{2}(P^{\mu\rho}P^{\nu\sigma}+P^{\mu\sigma}P^{\nu\rho})-\frac{1}{3}P^{\mu\nu}P^{\rho\sigma}\,,
}
with $P^{\mu\nu}$ defined by \cref{projection operator Pmunu}.

\subsection{Transverse vector}

The Hessian for the spin-1 gauge mode $\upsilon_{\mu}$ is given by
\al{
&\left(\Gamma_{(\upsilon\upsilon)}^{(2)}\right)^{\mu\nu}=
\frac{1}{\alpha}F\bar{\mathcal D}_1\left[  \bar{\mathcal D}_1+\alpha\frac{\bar R}{2} -\alpha \frac{U}{F}\right] P^{(v)\mu\nu}\,,
}
with $P^{(v)}$ the projection operator on the vector mode, $P^{(v)}{}_\mu{}^\mu=3$. Note that $\bar{\mathcal D}_1$ is defined in \cref{appeq: D1 and D0}.
For $\alpha\to0$ the inverse propagator for the gauge-vector mode becomes independent of $U$
\al{
\lim_{\alpha\to 0} \alpha\left( \Gamma_k^{(2)}\right)_{\upsilon\upsilon}^{\mu\nu}=F\left(\bar{\mathcal D}_1\right)^2P^{(v)\mu\nu}\,.
\label{spin 1 contribution in Landau gauge}
}
Note that the overall factor $1/\alpha$ cancels with that in the regulator function $\mathcal R_k^{\upsilon\upsilon}(\bar\Delta_V)$ within the flow equation.

\subsection{Scalars}

We next turn to the Hessian for the scalar modes.
In the $(\sigma,u,\varphi)$-field basis, we obtain
\al{
\Gamma_{(00)}^{(2)}=
\pmat{
\begin{matrix} \left(\Gamma_{(00)}^{(2)}\right)_\text{grav} \end{matrix} & \begin{matrix}  \displaystyle \frac{1}{2}\left[- F' \left(\bar \Delta_S+\frac{\bar R}{4} \right)+ U' \right]\bar\phi\\[20pt]
\displaystyle \frac{1}{4}\left(-F'\bar R +2U'  \right)\bar\phi \bar\Delta_{S}\end{matrix} \\[30pt]
 \begin{matrix} \displaystyle \displaystyle \frac{1}{2}\left[ -F' \left(\bar \Delta_S+\frac{\bar R}{4} \right)+ U' \right] \bar\phi &&&  \displaystyle \frac{1}{4}\left(-F'\bar R +2U' \right)\bar\phi \bar\Delta_{S} \end{matrix}
&~~~~~~~~
\displaystyle  K\bar\Delta_S+m^2(\bar\rho) -\frac{1}{2} \xi(\bar\rho)\,\bar R
}\,,
\label{spin 0 matrix}
}
where we define $m^2\fn{\bar\rho}=U'+2\bar\rho U''$ and $\xi\fn{\bar\rho}=F'+2\bar\rho F''$.
Here the spin-0 gravitational part is given by the following $2\times 2$ matrix:
\al{
&\left(\Gamma_{(00)}^{(2)}\right)_\text{grav}=
\nn[1ex]&\scriptsize\pmat{
\displaystyle -\frac{F}{6}\left[ \left(\bar \Delta_S -\frac{3}{8}\bar R\right) \bar \Delta_S -\frac{U}{2F}\left( \bar \Delta_S-\frac{\bar R}{2}\right)\right] \left( \bar \Delta_S-\frac{\bar R}{3}\right)^{-1}+\frac{(\beta+1)^2}{16\alpha}\bar \Delta_S
&
\displaystyle \frac{F}{4}\left( \frac{U}{F}-\frac{\bar R}{4}\right)\bar\Delta_{S} + \frac{(\beta +1)(\beta -3)}{16 \alpha} \left(\bar \Delta_S+\frac{\bar R}{\beta -3}\right)\bar\Delta_{S} \\[20pt]
\displaystyle \frac{F}{4}\left( \frac{U}{F}-\frac{\bar R}{4}\right)\bar\Delta_{S} + \frac{(\beta +1)(\beta -3)}{16 \alpha} \left(\bar \Delta_S+\frac{\bar R}{\beta -3}\right)\bar\Delta_{S}
&
\displaystyle \frac{F}{16}\left( \bar \Delta_S-\frac{\bar R}{2}\right)\left( \bar R-\frac{4U}{F}\right) \bar \Delta_S +\frac{(\beta -3)^2 }{16 \alpha } \left( \bar \Delta_S+\frac{\bar R}{\beta -3}\right)^2\bar \Delta_S
}\,.
\label{Hessian spin 0 part}
}

For general $\alpha$ and $\beta$ the Hessian in the scalar sector is rather complicated.
It simplifies considerably for a physical gauge fixing that corresponds to $\beta=-1$.
For $\beta=-1$ the factor $1/\alpha$ remains only in the Hessian for the spin-0 gauge mode $\Gamma^{(0)}_{(uu)}$.
In the limit $\alpha\to 0$ we can further neglect in the matrix \labelcref{spin 0 matrix} the elements mixing $u$ with $\sigma$ and $\varphi$.
They do not diverge for $\alpha\to0$ and drop out in the propagator that is the inverse of $\Gamma^{(2)}$.
As a result, for a physical gauge fixing the physical fluctuations and the gauge modes decouple.
The Hessian becomes block diagonal in physical and gauge modes.
For the physical modes one obtains the $2\times2$ matrix in the $(\sigma,\varphi)$ basis
%
\al{
&\left( \Gamma_{(00)}^{(2)}\right)_\text{ph}
=\pmat{
\displaystyle -\frac{F}{6}\left[ \left(\bar \Delta_S -\frac{3}{8}\bar R\right) \bar \Delta_S -\frac{U}{2F}\left( \bar \Delta_S -\frac{\bar R}{2}\right)\right] \left( \bar \Delta_S -\frac{\bar R}{3}\right)^{-1}
 &&  \displaystyle \frac{1}{2}\left[ - F' \left(\bar \Delta_S+\frac{\bar R}{4} \right)+ U' \right] \bar\phi \\[5ex]
\displaystyle \frac{1}{2}\left[- F' \left(\bar \Delta_S+\frac{\bar R}{4} \right)+ U' \right]\bar\phi && K\bar \Delta_S+m^2(\bar\rho) -\displaystyle\frac{1}{2} \xi(\bar\rho)\,\bar R
}\,,\label{physical spin 0 mode two point function}
}

In the basis with the redefinition of $\sigma$ as in \cref{app:eq: redefinition of sigma}, i.e., $(\tilde\sigma, \varphi)$, the Hessian is given by
\al{
&\left( \Gamma_{(00)}^{(2)}\right)_\text{ph}
=\pmat{
\displaystyle -\frac{F}{6}\left[ \left(\bar \Delta_S -\frac{3}{8}\bar R\right) \bar \Delta_S -\frac{U}{2F}\left( \bar \Delta_S -\frac{\bar R}{2}\right)\right] \left( \bar \Delta_S -\frac{\bar R}{3}\right)
 &&&  \displaystyle \frac{\bar\phi}{2}\left[ - F' \left(\bar \Delta_S+\frac{\bar R}{4} \right)+ U' \right] \left( \bar \Delta_S -\frac{\bar R}{3}\right)\\[5ex]
\displaystyle \frac{\bar\phi}{2}\left[- F' \left(\bar \Delta_S+\frac{\bar R}{4} \right)+ U' \right]\left( \bar \Delta_S -\frac{\bar R}{3}\right) &&& K\bar \Delta_S+m^2(\bar\rho) -\displaystyle\frac{1}{2} \xi(\bar\rho)\,\bar R
}\,,\label{new physical spin 0 mode two point function}
}

The Hessian for the gauge scalar mode reads
\al{
\left( \Gamma_{(00)}^{(2)}\right)_\text{gauge}=\Gamma_{(uu)}^{(2)} =\displaystyle\left(\bar \Delta_S-\frac{\bar R}{4}\right)^2\bar \Delta_S\,.
\label{measure spin 0 mode two point function}
}
%

\subsection{Ghost fields}

Finally, the Hessians for the ghost field obtains from \cref{ghostaction} as
\al{ 
&\left(\Gamma_{(\bar C^\perp C^\perp)}^{(2)}\right)^{\mu\nu}=\bar{\mathcal D}_1\, P^{(v)\mu\nu}\,,&
&\Gamma_{(\bar CC)}^{(2)}=\left( \bar \Delta_S +\frac{\bar R}{\beta-3}\right)\bar \Delta_S\,.
\label{ghost Hessians}
}
Note that $\bar{\mathcal D}_1$ is defined in \cref{appeq: D1 and D0}.

\section{Heat kernel evaluation of the flow generator}
\label{heat kernel methods}

Let us briefly summarize the the heat kernel formulas used in the gravitational setting. Suppose that the flow generator $\zeta_k$ takes
\al{
\zeta_k=\sum_i a_i\,\text{tr}\,W\fn{\bar\Delta_i}\,,
\label{appeq: flow kernels}
}
with $\bar\Delta_i=-\bar g^{\mu\nu}\bar D_\mu \bar D_\nu- {\bf U}$ appropriate differential operators acting on $i$ fields such as spin-2 traceless-transverse tensor, etc., and $a_i$ weight factors. 
The right-hand side of \cref{appeq: flow kernels} is 
\al{
\text{tr}\,W\fn{\bar\Delta}&=\sum_m W\fn{\lambda_m}= \sum_m \int^\infty_0\df z\,\delta\fn{z-\lambda_m}\,W\fn{z}
=\int^\infty_0\df z\,W\fn{z}\int^{\gamma+i\infty}_{\gamma-i\infty} \frac{\df s}{2\pi i}e^{sz}\,\text{tr}\,e^{-s\bar\Delta}\,,
\label{appeq: flow kernel 2}
}
where we use the identity $\sum_m \exp\fn{-s\lambda_m}=\text{tr}\,\exp\fn{-s\bar\Delta}$. 

Next, one expands the trace of $e^{-s\bar\Delta}$ in terms of $s$ such that
\al{
\text{tr}\,e^{-s\bar\Delta}
=\frac{1}{16\pi^2}\int_x\sqrt{\bar g} \left\{ \text{tr} [c_0\fn{\bar\Delta}] s^{-2}+ \text{tr}[c_2 \fn{\bar\Delta}] s^{-1}+ \text{tr}[ c_4\fn{\bar\Delta}] +\cdots\right\}\,.
\label{appeq: heat kernel expansion}
}
The coefficients $\text{tr} [c_n\fn{\bar\Delta}]$ are well known as the heat kernel coefficients. Inserting \cref{appeq: heat kernel expansion} into \cref{appeq: flow kernel 2} yields
\al{
\text{tr}\,W\fn{\bar\Delta}
=\frac{1}{16\pi^2}\sum_{n=0}^\infty Q_{2-n}\int_x\sqrt{\bar g}\,\text{tr}[c_{2n}\fn{\bar\Delta}]\,,
}
where we have performed the inverse Laplace transformation (Mellin's inverse formula) such that
\al{
Q_n=\int^\infty_0\df z\,W\fn{z}\int^{\gamma +i\infty}_{\gamma-i\infty}\frac{\df s}{2\pi i} e^{sz}s^{-n}
=\frac{1}{\Gamma\fn{n}}\int^\infty_0 \df z\,z^{n-1}W\fn{z} \qquad (n\geq 1)\,,
\label{Q-function}
}
with $\Gamma(n)$  being the Gamma function.
For the first three terms ($n=0$, $1$, $2$), \cref{Q-function} becomes
\al{
&Q_2=\int^\infty_0\df z\,zW\fn{z}\,,&
&Q_1=\int^\infty_0\df z\,W\fn{z}\,,&
&Q_0=W\fn{z=0}\,,
}
For negative $n$, one has
\al{
Q_{-n}=(-1)^n\frac{\p^n W}{\p z^n}\Bigg|_{z=0}\,,\qquad n\geq 0\,.
}
For $[\bar\Delta, {\bf U}]=0$, the heat kernel coefficients are~\cite{Gilkey:1995mj}
\al{
&c_0(\bar\Delta) = {\bf 1},\\
&c_2(\bar\Delta) = \frac{\bar R}{6} {\bf 1} - {\bf U},\\
&c_4(\bar\Delta) = 
 \left( \frac{1}{180}\bar R_{\mu\nu\rho\sigma}^2 -\frac{1}{180}\bar R_{\mu\nu}^2 + \frac{1}{72}\bar R^2
 -\frac{1}{30}\bar{\square} \bar R \right){\bf 1}
 +\frac{1}{2}{\bf U}^2 + \frac{1}{6}\bar{\square} {\bf U} + \frac{1}{12}\Omega_{\mu\nu}\Omega^{\mu\nu}
- \frac{\bar R}{6}{\bf U}\,,
}

In this work, it is sufficient to take $\text{tr}[c_0]$ and $\text{tr}[c_2]$ which are given by
\al{
&\text{tr}[c_0]=b_0\,,&
&\text{tr}[c_2]=b_2 \bar R \,.
}
We list the value of $b_0$ and $b_2$ for $\bar\Delta_i$ with ${\bf U}=0$ acting on several fields in \Cref{hkcs}.

\begin{table*}
\begin{center}
\caption{Heat kernel coefficients for the individual fields in maximally symmetric four dimensional spacetime. ``T'' and ``TT'' denote ``transverse'' and ``transverse-traceless,'' respectively.}
\label{hkcs}
\begin{tabular}{|c|c|c|c|c|c|}
\hline
    \makebox[1cm]{}  &  \makebox[1.1cm]{tensor} & \makebox[1.5cm]{T-tensor} & \makebox[1.7cm]{TT-tensor}  & \makebox[1.5cm]{T-vector}  & \makebox[1.1cm]{scalar} \\
      & ($h_{\mu\nu}$) & ($f_{\mu\nu}$) & $(t_{\mu\nu})$ & ($A_\mu^\text{T}$, $\upsilon_\mu$, $C^\perp_\mu$) & ($u$, $\sigma$, $\varphi$, $C$) \\
\hline
& & &  & & \\
$b_0$ & $10$ & $9$ & $5$ & $3$ & $1$ \\
& & & &  & \\
\hline
& & & &  & \\
$b_2$ & $\displaystyle \frac{5}{3}$ & $\displaystyle \frac{3}{2}$ & $\displaystyle -\frac{5}{6}$  &$\displaystyle \frac{1}{4}$ & $\displaystyle \frac{1}{6}$ \\
& & & &  & \\
\hline
\end{tabular}
\end{center}
\end{table*}

\section{Flow generator}
\label{evaluation of flow equations}

In this appendix, we calculate the flow equations.
In the current setup, the flow generator consists of three parts
\al{
\p_t \Gamma_k= \pi_2+\pi_0 + \eta_k\,.
}
Here, $\pi_2$ denotes the contribution from fluctuations of the TT-tensor field $t_{\mu\nu}$, while $\pi_0$ involves both the scalar field $\varphi$ and the physical scalar mode in the metric field.
These contributions correspond to the physical modes, while the measure modes such as the ghost fields and the vector and unphysical scalar modes in the metric field are collected in $\eta_k$.

\subsection{Physical metric fluctuations}

We first evaluate the contributions from the physical fluctuations.
There are the TT mode ($t_{\mu\nu}$) and the physical spin-0 modes ($\varphi$ and $\tilde\sigma$), whose forms of the flow generators are given by
\al{
&\pi_2=\frac{1}{2}\text{Tr}_{(2)}\frac{\p_t \mathcal R_k}{\Gamma_k^{(2)}+\mathcal R_k}\Bigg|_{tt}\,,
\label{physical mode generators tt}
\\
&\pi_0=\frac{1}{2}\text{Tr}_{(0)}\frac{\p_t \mathcal R_k}{\Gamma_{k}^{(2)}+\mathcal R_k}\Bigg|_\text{ph} + J_{\tilde\sigma}\,,
\label{physical mode generators sigmasigma}
}
respectively.
The two-point functions $\Gamma^{(2)}_k$ for the TT mode and the physical spin-0 modes are shown in \cref{TT mode two point function,physical spin 0 mode two point function}, respectively. The Jacobian $J_{\tilde\sigma}$ is given in \cref{app eq: Jacobian for sigma tilde}.

To extract the explicit form of the beta functions, we define the dimensionless quantities
\al{
&\tilde \rho=\frac{\bar\phi^2}{2k^2}=\frac{\bar\rho}{k^2}\,,\qquad
u\fn{\tilde\rho}=\frac{U\fn{\bar\rho}}{k^4}\,,\qquad
w\fn{\tilde \rho}=\frac{F\fn{\bar\rho}}{2k^2}\,,\qquad
\kappa\fn{\tilde \rho}= K\fn{\bar\rho}\,,
\nn[2ex]
&\tilde\xi(\tilde\rho)= \xi(\bar\rho)=F'+2\bar\rho F''=2w'+4\tilde \rho w''\,,\qquad
\tilde m^2\fn{\tilde \rho}=\frac{m^2(\bar\rho)}{k^2}=\frac{U'+2\bar\rho U''}{k^2}=u'+2\tilde \rho u''\,.
}
Primes acting on the dimensionless quantities denote the derivative with respect to $\tilde\rho$. Hereafter we introduce $\tau$ derivative as $\p_\tau=-2\tilde\rho\p_{\tilde\rho}$.

\subsubsection{Spin-2 TT mode}

We can evaluate the contributions \labelcref{physical mode generators tt}.
For the TT mode, our choice of IR cutoff function corresponds to the type-II cutoff function defined in Ref.~\cite{Codello:2008vh}, with
\al{
W_2\fn{z}=\frac{\p_t ( wk^2 R_k\fn{z})}{\left(wk^2 P_k\fn{z} - uk^4\right)}\,.
}
The heat kernel method yields the flow generator for the TT mode fluctuation as
\al{
\pi_2&=\frac{\Omega}{(4\pi)^2}\bigg[ \frac{10}{3} k^4\frac{w + \frac{\p_t w + \p_\tau w}{8} }{w- u}  
- \frac{25}{4} k^2\frac{w+ \frac{\p_t w + \p_\tau w}{6} }{w- u} \bar R \bigg]\,,
}
where $\Omega=\int_x\sqrt{\bar g}$ is the spacetime volume where we use
\al{
\frac{\p_t (w(\tilde\rho)k^2)}{k^2}= 2 w(\tilde\rho) + \p_t w(\tilde\rho) - 2\tilde\rho\p_{\tilde\rho} w(\tilde\rho)
= 2 w(\tilde\rho) + \p_t w(\tilde\rho) + \p_\tau w(\tilde\rho)\,.
}
%

\subsubsection{Spin-0 physical modes}

We now calculate the contributions from the spin-0 physical scalar fields. Within the current truncation, it suffices to deal with terms up to linear order in $\bar R$.
\al{
&\Gamma_k^{\tilde\sigma\tilde\sigma}\fn{z}
= -\frac{1}{3}\left[ \left( \frac{F}{2} z^3 -  \frac{U}{4}z^2\right) - \frac{\bar R}{24}\left( \frac{17F}{2}z^2 - 5U^2 z\right) + \mathcal O(\bar R^2) \right]\,,\\
&\Gamma_k^{\varphi\varphi}\fn{z}= K(\bar\rho) z + m^2(\bar\rho) -\frac{1}{2} \xi(\bar\rho)\,\bar R\,,\\
&\Gamma_k^{\tilde\sigma\varphi}(z) =\Gamma_k^{\varphi\tilde\sigma}(z)
=   - \bar\phi \left[ \left( \frac{F'}{2} z^2 - \frac{U'}{2}z \right) - \frac{1}{12}\left( \frac{F'}{2}z - 2U' \right)\bar R + \mathcal O(\bar R^2) \right]\,.
}
Hereafter we neglect higher operator terms $\mathcal O(\bar R^2)$. We employ the cutoff function for the physical scalar modes as
\al{
\label{appeq:regulator}
{\mathcal R}_k^\text{ph}\fn{z}
=\pmat{
{\mathcal R}_k^{\tilde\sigma\tilde\sigma}(z)  && {\mathcal R}_k^{\tilde\sigma\varphi}(z)   \\[3ex]
{\mathcal R}_k^{\varphi\tilde\sigma}(z) &&  {\mathcal R}_k^{\varphi\varphi}(z) 
}
=\pmat{
\Gamma_k^{\tilde\sigma\tilde\sigma}\fn{P_k}-\Gamma_k^{\tilde\sigma\tilde\sigma}\fn{z}  && \Gamma_k^{\tilde\sigma\varphi}\fn{P_k}-\Gamma_k^{\tilde\sigma\varphi}\fn{z}\\[3ex]
\Gamma_k^{\varphi\tilde\sigma}\fn{P_k}-\Gamma_k^{\varphi\tilde\sigma}\fn{z}  &&  \Gamma_k^{\varphi\varphi}\fn{P_k}-\Gamma_k^{\varphi\varphi}\fn{z}
}\Bigg|_{\bar R\to 0}\,.
}
For this cutoff function, the Laplacian $z=\bar\Delta_S$ only in the the curvature-scalar independent terms is replaced with $P_k\fn{z}=z+R_k\fn{z}$.
The full two-point function for the physical scalar modes reads
\al{
\fn{\Gamma_{k}^{(2)}+\mathcal R_k}\Big|_\text{ph}
&=\pmat{
 -\frac{1}{3}\left( \frac{F}{2} P_k^3 -  \frac{U}{4}P_k^2\right) + \frac{\bar R}{72}\left( \frac{17F}{2}z^2 - 5U^2 \right) &&  - \bar\phi \left[ \left( \frac{F'}{2} P_k^2 - \frac{U'}{2}P_k \right) - \frac{1}{12}\left( \frac{F'}{2}z - 2U' \right)\bar R \right] \\[3ex]
- \bar\phi \left[ \left( \frac{F'}{2} P_k^2 - \frac{U'}{2}P_k \right) - \frac{1}{12}\left( \frac{F'}{2}z - 2U' \right)\bar R  \right] 
&& K P_k + m^2-\frac{1}{2} \xi\,\bar R
}\nn
&=
\pmat{
 -\frac{1}{3}\left( \frac{F}{2} P_k^3 -  \frac{U}{4}P_k^2\right) &&  - \bar\phi \left( \frac{F'}{2} P_k^2 - \frac{U'}{2}P_k \right)  \\[3ex]
- \bar\phi  \left( \frac{F'}{2} P_k^2 - \frac{U'}{2}P_k \right)  
&& K P_k + m^2
}
+ \pmat{
 \frac{1}{72}\left( \frac{17F}{2}z^2 - 5U^2 \right) && \frac{\bar\phi }{12}\left( \frac{F'}{2}z - 2U' \right) \\[3ex]
\frac{\bar\phi }{12}\left( \frac{F'}{2}z - 2U' \right)
&& -\frac{1}{2} \xi
}\bar R\nn
&\equiv G_{\rm ph}^{-1} + I_{\rm ph}\bar R\,.
\label{appeq: modified propagator}
}
Here, the propagator matrix for the physical scalar fields is given by
\al{
 G_{\rm ph} &= \frac{1}{ -\frac{1}{3}\left( \frac{F}{2} P_k^3 -  \frac{U}{4}P_k^2\right)\left(  K P_k + m^2 \right) - 2\bar\rho\left( \frac{F'}{2} P_k^2 - \frac{U'}{2}P_k \right)^2 }
 \pmat{
K P_k + m^2 &&   \bar\phi \left( \frac{F'}{2} P_k^2 - \frac{U'}{2}P_k \right)  \\[3ex]
\bar\phi  \left( \frac{F'}{2} P_k^2 - \frac{U'}{2}P_k \right)  &&  -\frac{1}{3}\left( \frac{F}{2} P_k^3 -  \frac{U}{4}P_k^2\right) 
}\nn
&= \frac{1}{\left( \frac{F}{2} P_k^3 -  \frac{U}{4}P_k^2\right)\left(  K P_k + m^2 \right) + 6\bar\rho\left( \frac{F'}{2} P_k^2 - \frac{U'}{2}P_k \right)^2 }
 \pmat{
-3(K P_k + m^2) &&   -3\bar\phi \left( \frac{F'}{2} P_k^2 - \frac{U'}{2}P_k \right)  \\[3ex]
-3\bar\phi  \left( \frac{F'}{2} P_k^2 - \frac{U'}{2}P_k \right)  &&   \frac{F}{2} P_k^3 -  \frac{U}{4}P_k^2 }
\nn
&\equiv \frac{1}{\left( \frac{F}{2} P_k^3 -  \frac{U}{4}P_k^2\right)\left(  K P_k + m^2 \right) + 6\bar\rho\left( \frac{F'}{2} P_k^2 - \frac{U'}{2}P_k \right)^2 } \widetilde G_\text{ph}
\nn[2ex]
&\equiv
 \pmat{
\mathcal P_{\tilde\sigma\tilde\sigma} &&  \mathcal P_{\tilde\sigma\varphi} \\[2ex]
\mathcal P_{\varphi\tilde\sigma}  &&   \mathcal P_{\varphi\varphi}
}
\,.
\label{appeq: propagator matrix}
}
Note that
\al{
&\p_t {\mathcal R}_k^\text{ph}\fn{z}
=\pmat{
 -\frac{1}{3}\p_t\left( wk^2(P_k^3 -z^3) -  \frac{uk^4}{4}(P_k^2-z^2)\right) &&  - \bar\phi \p_t\left( w' (P_k^2 -z^2)- \frac{u'k^2}{2}(P_k-z) \right)  \\[3ex]
- \bar\phi  \p_t\left( w' (P_k^2-z^2) - \frac{u'k^2}{2}(P_k-z) \right)  
&& \p_t(\kappa (P_k -z))}\,.
\label{appeq: regulator function with derivative}
}

The inverse form of \cref{appeq: modified propagator} is then computed in polynomials of $\bar R$ as
\al{
\fn{\Gamma_{k}^{(2)}+\mathcal R_k}\Big|_\text{ph}^{-1}
= (G_{\rm ph}^{-1} + I_{\rm ph} \bar R)^{-1}
=(1 + G_{\rm ph} I_{\rm ph}  \bar R)^{-1}G_{\rm ph}
=G_{\rm ph} - G_{\rm ph} I_{\rm ph} G_{\rm ph} \bar R+\cdots\,,
}
from which the flow kernel becomes
\al{
\frac{1}{2}\text{Tr}_{(0)}\frac{\p_t \mathcal R_k}{\Gamma_{k}^{(2)}+\mathcal R_k}\Bigg|_\text{ph}
=\frac{1}{2}\Tr_{(0)} \left[G_{\rm ph}\p_t \mathcal R^{\rm ph}_k \right] - \frac{1}{2}\Tr_{(0)} \left[ G_{\rm ph} I_{\rm ph} G_{\rm ph} \p_t \mathcal R^{\rm ph}_k \right]\bar R + \cdots\,.
}
Here, the first term in the right-hand side is
\al{
\frac{1}{2}\Tr_{(0)}[G_{\rm ph}\p_t \mathcal R^{\rm ph}_k] 
&= \frac{\Omega}{2(4\pi)^2} b^{(0)}_{0} Q_{2}[G_{\rm ph}(z)\p_t \mathcal R^{\rm ph}_k(z)]
+ \frac{\Omega}{2(4\pi)^2} b^{(0)}_{2}\bar R\, Q_{1}[G_{\rm ph}(z)\p_t \mathcal R^{\rm ph}_k(z)]\,,\\[2ex]
- \frac{1}{2}\Tr_{(0)}[ G_{\rm ph} I_{\rm ph} G_{\rm ph} \p_t \mathcal R^{\rm ph}_k ]\bar R
&= -\frac{\Omega}{2(4\pi)^2} b_0^{(0)}Q_2[G_{\rm ph}(z) I_{\rm ph}(z) G_{\rm ph}(z) \p_t \mathcal R^{\rm ph}_k(z)]\bar R\,.
}

Finally, we evaluate the contribution from the Jacobian \labelcref{app eq: Jacobian for sigma tilde} for $\tilde\sigma$ which evaluates to
\al{
J_{\tilde\sigma} &= -\frac{1}{2}\Tr_{(0)}\frac{\p_t \mathcal R_k(\bar{\mathcal D}_0)}{P_k(\bar{\mathcal D}_0 )}
-\frac{1}{2}\Tr_{(0)}\frac{\p_t \mathcal R_k(\bar\Delta_S -\frac{\bar R}{3})}{P_k(\bar\Delta_S -\frac{\bar R}{3} )}
=-\frac{\Omega}{2(4\pi)^2}\left( k^4 +\frac{5}{6}k^2 \bar R \right)\,.
}
To summarize, the flow equations for $U$ and $F$ are
\al{
\pi_0\Big|_{\bar R^0}&=\frac{\Omega}{2(4\pi)^2} b^{(0)}_{0} Q_{2}[G_{\rm ph}(z)\p_t \mathcal R^{\rm ph}_k(z)] -\frac{\Omega k^4}{2(4\pi)^2}\,,\\
\pi_0\Big|_{\bar R^1}&=\frac{\Omega}{2(4\pi)^2} b^{(0)}_{2} Q_{1}[G_{\rm ph}(z)\p_t \mathcal R^{\rm ph}_k(z)]  -\frac{\Omega}{2(4\pi)^2} b_0^{(0)}Q_2[G_{\rm ph}(z) I_{\rm ph}(z) G_{\rm ph}(z) \p_t \mathcal R^{\rm ph}_k(z)] -\frac{\Omega k^2}{2(4\pi)^2}\frac{5}{6}\,.
}
%

\subsection{Measure contribution}
\label{Measure contributions}

We now calculate the measure contribution $\eta_k$. This contribution takes the simple form~\cite{Wetterich:2016ewc}
\al{
\eta_k=-\frac{1}{2}\text{Tr}_{(1)}\frac{\p_t P_k\fn{\bar{\mathcal D}_1}}{P_k\fn{\bar{\mathcal D}_1}}-\frac{1}{2}\text{Tr}_{(0)}\frac{\p_t P_k\fn{\bar{\mathcal D}_0}}{P_k\fn{\bar{\mathcal D}_0}}\,,
\label{measure contributions}
}
with 
\al{
&\bar{\mathcal D}_1=\bar \Delta_V-\frac{\bar R}{4}\,,&
&\bar{\mathcal D}_0=\bar \Delta_S-\frac{\bar R}{4}\,.
}
To see this, we evaluate each of the contributions from measure modes which arise from the gauge fluctuations, the ghost fluctuations, and the regularization of the Jacobian.

\subsubsection{Spin-1 measure modes}

The spin-1 measure contribution is given by
\al{
\eta_1=\delta^{(1)}_k-\epsilon^{(1)}_k\,,
\label{eta_1k}
}
with the spin-1 vector gauge mode \labelcref{spin 1 contribution in Landau gauge} and the Jacobians~\labelcref{app: eq: kappa Jacobian} for the spin-1 gauge field
\al{
\delta^{(1)}_k &= \lim_{\alpha\to 0}\frac{1}{2}\text{Tr}_{(1)}\frac{\p_t \mathcal R_k}{\Gamma_k^{(2)}+\mathcal R_k}\Bigg|_{\upsilon\upsilon} +J_{\upsilon}
=\text{Tr}_\text{(1T)}\frac{\p_t  P_k\fn{\bar{\mathcal D}_1}}{P_k\fn{\bar{\mathcal D}_1}} -\frac{1}{2}\text{Tr}_\text{(1T)}\frac{\p_t  P_k\fn{\bar{\mathcal D}_1}}{P_k\fn{\bar{\mathcal D}_1}}
=\frac{1}{2}\text{Tr}_\text{(1T)}\frac{\p_t  P_k\fn{\bar{\mathcal D}_1}}{P_k\fn{\bar{\mathcal D}_1}}\,,
}
and the spin-1 ghost mode \labelcref{ghost Hessians}
\al{
\epsilon^{(1)}_k &=\text{Tr}_\text{(1T)}\frac{\p_t \mathcal R_k}{\Gamma_k^{(2)}+\mathcal R_k}\Bigg|_{\bar C^\perp C^\perp}
=\text{Tr}_\text{(1T)}\frac{\p_t  P_k\fn{\bar{\mathcal D}_1}}{P_k\fn{\bar{\mathcal D}_1}}\,.
}
This sums up to the measure contribution from the spin-1 modes
\al{
\eta_1= -\frac{1}{2}\text{Tr}_\text{(1T)}\frac{\p_t  P_k\fn{\bar{\mathcal D}_1}}{P_k\fn{\bar{\mathcal D}_1}}
= -\frac{1}{2(4\pi)^2}\int_x\sqrt{\bar g}\left[ 3k^4  + 2 k^2\bar R \right]\,.
\label{app: eq: eta1}
}
%

\subsubsection{Spin-0 measure modes}

Next we discuss the spin-0 measure contribution.
The spin-0 gauge mode coming from the metric fluctuation \labelcref{measure spin 0 mode two point function} and the corresponding Jacobian \labelcref{app eq: Jacobian for u}  is
\al{
\delta^{(0)}_k &= \lim_{\alpha\to 0}\frac{1}{2}\text{Tr}_{(0)}\frac{\p_t \mathcal R_k}{\Gamma_k^{(2)}+\mathcal R_k}\Bigg|_{uu} + J_{u}\nn
&=\left( \text{Tr}_{(0)}\frac{\p_t  P_k\fn{\bar{\mathcal D}_0}}{P_k\fn{\bar{\mathcal D}_0}} 
+\frac{1}{2}\text{Tr}_{(0)}\frac{\p_t  P_k\fn{\bar{\Delta}_S}}{P_k\fn{\bar{\Delta}_S}} \right)
 -\frac{1}{2}\left( \text{Tr}_{(0)}\frac{\p_t  P_k\fn{\bar{\mathcal D}_0}}{P_k\fn{\bar{\mathcal D}_0}}
 + \text{Tr}_{(0)}\frac{\p_t  P_k\fn{\bar{\Delta}_S}}{P_k\fn{\bar{\Delta}_S}}\right) \nn[1ex]
&=\frac{1}{2}\text{Tr}_{(0)}\frac{\p_t  P_k\fn{\bar{\mathcal D}_0}}{P_k\fn{\bar{\mathcal D}_0}}\,,
}
while the spin-0 ghost mode \labelcref{ghost Hessians} is
\al{
-\epsilon^{(0)}_k &=-\text{Tr}_{(0)}\frac{\p_t \mathcal R_k}{\Gamma_k^{(2)}+\mathcal R_k}\Bigg|_{\bar C C}
+ J_\text{gh}
=-\left(\text{Tr}_{(0)}\frac{\p_t P_k\fn{\bar{\mathcal D}_0}}{P_k\fn{\bar{\mathcal D}_0}} + \text{Tr}_{(0)}\frac{\p_t P_k\fn{\bar\Delta_S}}{P_k\fn{\bar \Delta_S}}\right) 
+ \text{Tr}_{(0)}\frac{\p_t P_k\fn{\bar \Delta_S}}{P_k\fn{\bar\Delta_S}}\nn
&=-\text{Tr}_{(0)}\frac{\p_t P_k\fn{\bar{\mathcal D}_0}}{P_k\fn{\bar{\mathcal D}_0}}\,,
}
where $J_\text{gh}$ is given in \cref{app eq: Jacobian for ghost}.
Thus, the measure contribution of the spin-0 measure modes becomes
\al{
\eta_0&=\delta^{(0)}_k-\epsilon^{(0)}_k= -\frac{1}{2}\text{Tr}_{(0)}\frac{\p_t  P_k\fn{\bar{\mathcal D}_0}}{P_k\fn{\bar{\mathcal D}_0}}
= -\frac{1}{2(4\pi)^2}\int_x\sqrt{\bar g}\left[ k^4  + \frac{5}{6} k^2\bar R \right]\,.
}
Together with \cref{app: eq: eta1}, the contributions from the measure modes are given by
\al{
\eta_k = \eta_1 + \eta_0
=-\frac{1}{2}\text{Tr}_\text{(1T)}\frac{\p_t  P_k\fn{\bar{\mathcal D}_1}}{P_k\fn{\bar{\mathcal D}_1}} -\frac{1}{2}\text{Tr}_{(0)}\frac{\p_t  P_k\fn{\bar{\mathcal D}_0}}{P_k\fn{\bar{\mathcal D}_0}}
=-\frac{1}{2(4\pi)^2}\int_x \sqrt{\bar g}\left(4 k^4 +\frac{17}{6}k^2 \bar R \right)\,.
}
%

\section{Quantum corrections to scalar kinetic term}
\label{eq: Quantum corrections to scalar kinetic term}

In this section, we show the detailed derivation of the flow equation for $K$ which is the coefficient of the kinetic term of the scalar field. 
For a derivation in a simple scalar field theory, see e.g., Refs.~\cite{Morris:1994ie,Berges:2000ew} for the derivation.

In the kinetial sector, background curvature operators such as $\bar R$ and $\bar R_{\mu\nu}$ do not appear. We therefore work on a flat background $\bar g_{\mu\nu}=\delta_{\mu\nu}$ for which $\sqrt{\bar g}=1$, $\bar R=0$ and $\tilde\sigma =(\bar\Delta_S)^{-1}\sigma=(-\p^2)^{-1}\sigma$. In this case, the kinetic term for the scalar field is
\al{
\label{eq:kineticterm}
\Gamma^{\rm kin}_k[\phi] = \frac{1}{2}\int_x K(\bar\rho) \delta^{\mu\nu}\p_\mu \bar\phi \p_\nu \bar\phi
\equiv \frac{1}{2}\int_x K(\bar\rho) \mathcal O_K\,.
}
From the kinetic term \labelcref{eq:kineticterm}, we obtain the flow of the kinetial $K$ as
\al{
\frac{\p_t K (\bar\rho)}{2}= \frac{1}{\Omega}\frac{\delta}{\delta \mathcal O_K}\p_t \Gamma_k\,,
\label{appeq: kinetial flow}
}
where $\Omega=\int_x=(2\pi)^4\delta^4(p=0)$ is the four dimensional spacetime volume.

Here, thanks to the properties of the metric fluctuations and tensorial structures of variations, we can reduce the computation of the flow kernel of $K$.
First, since we are interested in the corrections to the $p^2$ term for the two-point functions of $\bar\phi$, we symmetrize external momenta according to $\p_\mu \p_\nu \to \frac{1}{4}\;p^2\delta_{\mu\nu}$. Then, the contraction between the Kronecker delta and $H_2^{\mu\nu}$ becomes
\al{
\delta_{\mu\nu}H_2^{\mu\nu}
=\delta_{\mu\nu}\left[  \left( -\frac{1}{2}h^{\alpha\beta}h_{\alpha\beta} +\frac{1}{4} h^2\right)\delta^{\mu\nu} + 2h^\mu{}_\lambda h^{\nu\lambda}  - hh^{\mu\nu}\right]
=0\,.
\label{eqapp:trace of H2}
}
This implies that the tadpole diagrams arising from the metric fluctuations do not induce quantum corrections to the flow kernel of $K$.
Second, the threshold functions in the tadpole diagram do not carry the external momentum $p_\mu$, and therefore the loop effects arising from the four-point vertices without momentum dependence do not give contributions to the flow kernel of $K$.
Here, we have
\al{
H_1^{\mu\nu} &= \frac{1}{2} h\bar g^{\mu\nu} - h^{\mu\nu}
 = \frac{1}{2}\delta^{\mu\nu}\bar\Delta_S\tilde\sigma  - \frac{1}{3} (\delta^{\mu\nu} \bar\Delta_S - \p^\mu \p^\nu)\tilde\sigma + \cdots
 = \frac{1}{6}\delta^{\mu\nu}\bar\Delta_S\tilde\sigma + \frac{1}{3}\p^\mu \p^\nu \tilde\sigma + \cdots\,.
}
The relevant functional derivatives of the scalar kinetic term are
\al{
&\frac{\delta^2}{\delta\phi(x)\delta\phi(y)}\int_z\sqrt{g} \frac{K(\rho)}{2}g^{\mu\nu}\p^z_\mu \phi \p^z_\nu \phi\bigg|_{g=\delta, \phi=\bar\phi}
\nn
&=\bigg[
- K \p^2
-\frac{1}{2} (K'+ 2\bar\rho K'') \mathcal O_K
- (\bar\phi K')\p^2 \bar\phi 
- (\bar\phi K') \p_\mu\bar\phi \p^{\mu}  
\bigg]  \delta(x-y)\,,
\\[2ex]
&\frac{\delta^2}{\delta \tilde\sigma(x) \delta\phi(y)}
\int_z \frac{K(\rho)}{2} \left( \frac{1}{6}\delta^{\mu\nu}\bar\Delta_S^z\tilde\sigma + \frac{1}{3}\p^z{}^\mu \p^z{}^\nu \tilde\sigma\right) \p_\mu^z\phi \p_\nu^z \phi\bigg|_{g=\delta, \phi=\bar\phi}\nn
&=  \bigg[
\left( \frac{\bar\phi K'}{4} \mathcal O_K  + \frac{K}{3}\p^2\bar\phi  \right) \p^2 
+ \frac{K}{6} \p_\mu \bar\phi \p^2\p^{\mu} 
+ \left( \frac{\bar\phi K'}{6} (\p_\mu\bar\phi)(\p_\nu\bar\phi) 
 + \frac{K}{3}(\p_\mu\p_\nu \bar\phi) \right) \p^{\mu} \p^{\nu}+\cdots
\bigg] \delta(x-y)\,,
\\[2ex]
&\frac{\delta^2}{\delta\phi(x) \delta \tilde\sigma(y)}
\int_z \frac{K(\rho)}{2} \left( \frac{1}{6}\delta^{\mu\nu}\bar\Delta_S^z\tilde\sigma + \frac{1}{3}\p^z{}^\mu \p^z{}^\nu \tilde\sigma\right) \p_\mu^z\phi \p_\nu^z \phi\bigg|_{g=\delta, \phi=\bar\phi}\nn
&=\bigg[
\left( \frac{\bar\phi K'}{12}\mathcal O_K +\frac{K}{6} (\p^2\bar\phi) \right) \p^2
- \frac{K}{6}  \p_\mu \bar\phi\p^2\p^{\mu}  
- \left( \frac{\bar\phi K'}{6} \p_\mu \bar\phi \p_\nu \bar\phi + \frac{K}{3}(\p_\mu\p_\nu\bar\phi) \right) \p^\mu \p^\nu
  + \cdots\bigg] \delta(x-y)\,,
}
where $\cdots$ includes operators irrelevant for the computation of the flow equation for the kinetial.  
The term $\sim \frac{\delta^2 }{\delta \tilde\sigma\delta \tilde\sigma}$ acting on the scalar kinetic term does not contribute due to the property~\labelcref{eqapp:trace of H2}.
Thus the Hessian in momentum space for computing the flow equation for $K$ is given by
\al{
\left( \Gamma_k^{(2)} + \mathcal R_k\right)_\text{ph}
&= 
\pmat{
\displaystyle -\frac{1}{6}\left( FP_k -\frac{U}{2}\right)P_k^2
 &&&  \displaystyle \frac{\bar\phi}{2}\left( - F'P_k + U' \right) P_k  \\[3ex]
\displaystyle \frac{\bar\phi}{2}\left( - F'P_k + U' \right) P_k    &&& KP_k +m^2(\bar\rho) 
}\nn
&\quad
+
\pmat{
\displaystyle 0
 &&&  \displaystyle\mathcal X \p^2 + \mathcal Y^\mu \p^2 \p_\mu + \mathcal Z^{\mu\nu} \p_\mu \p_\nu \\[3ex]
\displaystyle \mathcal X' \p^2 - \mathcal Y^\nu \p^2 \p_\nu - \mathcal Z^{\mu\nu} \p_\mu \p_\nu  &&& \mathcal A^\mu \p_\mu + \mathcal B
}
= G_\text{ph}^{-1} + \mathcal V_\text{ph}\,.
}
where we have defined
\al{
\mathcal A^\mu &= - \bar\phi K' (\p^\mu \bar\phi)\,,\qquad
\mathcal B = -\frac{1}{2}(K'+2\bar\rho K'') \mathcal O_K   - \bar\phi K' (\p^2\bar\phi)\,,\nn[1ex]
\mathcal X &= \frac{\bar\phi K'}{4} \mathcal O_K+\frac{K}{3} \p^2\bar\phi \,,\qquad
\mathcal X' = \frac{\bar\phi K'}{12} \mathcal O_K+\frac{K}{6} \p^2\bar\phi \,,\nn
\mathcal Y^\mu &= \frac{K}{6}(\p^\mu\bar\phi),\qquad
\mathcal Z^{\mu\nu} = \frac{\bar\phi K'}{6}(\p^\mu\bar\phi)(\p^\nu\bar\phi) + \frac{K}{3}(\p^\mu \p^\nu \bar\phi)\,.
}

Next, we expand the full propagator into polynomials of the vertex operator $\mathcal V_\text{ph}$ as
\al{
\fn{\Gamma_{k}^{(2)}+\mathcal R_k}\Big|_\text{ph}^{-1}
= (G_\text{ph}^{-1} + \mathcal V_\text{ph})^{-1}
=(1 + G_{\rm ph} \mathcal V_\text{ph})^{-1}G_{\rm ph}
=G_{\rm ph} - G_{\rm ph} \mathcal V_\text{ph} G_{\rm ph} + G_{\rm ph} \mathcal V_\text{ph} G_{\rm ph}\mathcal V_\text{ph} G_{\rm ph}  +\cdots\,,
}
from which the flow equation for kinetial is obtained as
\al{
\label{eq:KinetialFlowterms}
&-\frac{1}{2} \Tr_{(0)}\left[ G_\text{ph}\mathcal V_\text{ph}  G_\text{ph}\p_t \mathcal R_k^\text{ph}  \right]
+\frac{1}{2} \Tr_{(0)}\left[ G_\text{ph}\mathcal V_\text{ph}  G_\text{ph}\mathcal V_\text{ph} G_\text{ph} \p_t \mathcal R_k^\text{ph}  \right]\nn
&\qquad
=-\frac{1}{2} \Tr_{(0)}\left[ G_\text{ph}\mathcal V^{(2)}_\text{ph}  G_\text{ph}\p_t \mathcal R_k^\text{ph}  \right]
+\frac{1}{2} \Tr_{(0)}\left[ G_\text{ph}\mathcal V^{(1)}_\text{ph}  G_\text{ph}\mathcal V^{(1)}_\text{ph} G_\text{ph} \p_t \mathcal R_k^\text{ph}  \right]+\cdots\,,
}
where we have defined 
\al{
&\mathcal V^{(2)}_\text{ph}
=\pmat{
\displaystyle 0
 &&&  \displaystyle\mathcal X \p^2 + \mathcal Z^{\mu\nu} \p_\mu \p_\nu \\[3ex]
\displaystyle \mathcal X' \p^2  - \mathcal Z^{\mu\nu} \p_\mu \p_\nu  &&& \mathcal B
}\,,&
&
\mathcal V^{(1)}_\text{ph}
=\pmat{
\displaystyle 0
 &&&  \mathcal Y^\mu \p^2 \p_\mu \\[3ex]
- \mathcal Y^\nu \p^2 \p_\nu  &&& \mathcal A^\mu \p_\mu
}\,.
}
The first term in \cref{eq:KinetialFlowterms} corresponds to the first diagram in \Cref{fig:diag}, while the second two diagrams involve in the second term in  \cref{eq:KinetialFlowterms}. 
The symmetrization $\p_\mu \p_\nu \to \delta_{\mu\nu}\p^2/4$ in the functional trace (Tr${}_{(0)}$) yields
\begin{subequations}
\label{appeq:symmetrized}
\al{
&\mathcal X \p^2 + \mathcal Z^{\mu\nu}\p_\mu \p_\nu 
\to \left( \frac{\bar\phi K'}{4} \mathcal O_K+\frac{K}{3} \p^2\bar\phi \right) \p^2 +\frac{1}{4} \left( \frac{\bar\phi K'}{6}\mathcal O_K + \frac{K}{3}\p^2 \bar\phi \right)\p^2
= \left( \frac{7}{24}\bar\phi K' \mathcal O_K + \frac{5K}{12} \p^2 \bar\phi  \right) \p^2
=(\mathcal V_\text{ph}^{(2)})^{12}\p^2
\,,
\\
&\mathcal X' \p^2 - \mathcal Z^{\mu\nu}\p_\mu \p_\nu \to \left( \frac{\bar\phi K'}{12} \mathcal O_K+\frac{K}{6} \p^2\bar\phi \right) \p^2 - \frac{1}{4} \left( \frac{\bar\phi K'}{6}\mathcal O_K + \frac{K}{3}\p^2 \bar\phi \right)\p^2
= \left( \frac{1}{24}\bar\phi K' \mathcal O_K + \frac{K}{12} \p^2 \bar\phi  \right) \p^2
=(\mathcal V_\text{ph}^{(2)})^{21}\p^2\,,
}
\end{subequations}
and
\al{
\label{appeq:symmetrized2}
&\mathcal A^\mu \mathcal A^\nu \p_\mu \p_\nu \to \frac{2 \bar\rho K'^2 }{4}\mathcal O_K \p^2\,,&
&\mathcal A^\mu \mathcal Y^\nu \p^2\p_\mu \p_\nu \to -\frac{\bar\phi K' K}{6\times 4} \mathcal O_K \p^4 \, ,&
&\mathcal Y^\mu \mathcal Y^\nu \p^4 \p_\mu \p_\nu \to \frac{K^2}{36\times 4} \mathcal O_K \p^6\,.
}
Evaluating \cref{eq:KinetialFlowterms} with the propagator \labelcref{appeq: regulator function with derivative}, the regulator \labelcref{appeq:regulator} and the vertices \labelcref{appeq:symmetrized,appeq:symmetrized2}, we have
\al{
&-\frac{1}{2} \Tr_{(0)}\left[ G_\text{ph}\mathcal V^{(2)}_\text{ph}  G_\text{ph}\p_t \mathcal R_k^\text{ph}  \right]
+\frac{1}{2} \Tr_{(0)}\left[ G_\text{ph}\mathcal V^{(1)}_\text{ph}  G_\text{ph}\mathcal V^{(1)}_\text{ph} G_\text{ph} \p_t \mathcal R_k^\text{ph}  \right]\nn
&=
-\frac{1}{2}\frac{1}{(4\pi)^2}\int_x {\rm tr}\Bigg[
{\scriptsize
\pmat{
\mathcal P_{\tilde\sigma\tilde\sigma} &  \mathcal P_{\tilde\sigma\varphi} \\[2ex]
\mathcal P_{\varphi\tilde\sigma}  &   \mathcal P_{\varphi\varphi}
}\pmat{
0 & (\mathcal V_\text{ph}^{(2)})^{12}(\p^2) \\[2ex]
(\mathcal V_\text{ph}^{(2)})^{21} (\p^2) & \mathcal B
}\pmat{
\mathcal P_{\tilde\sigma\tilde\sigma} &  \mathcal P_{\tilde\sigma\varphi} \\[2ex]
\mathcal P_{\varphi\tilde\sigma}  &   \mathcal P_{\varphi\varphi}
}\pmat{
\p_t{\mathcal R}_k^{\tilde\sigma\tilde\sigma}  & \p_t{\mathcal R}_k^{\tilde\sigma\varphi}   \\[2ex]
\p_t{\mathcal R}_k^{\varphi\tilde\sigma} &  \p_t{\mathcal R}_k^{\varphi\varphi}
}
}
\Bigg]\nn
&
+\frac{1}{2}\frac{1}{(4\pi)^2}\int_x{\rm tr}\Bigg[
{\scriptsize
\pmat{
\mathcal P_{\tilde\sigma\tilde\sigma} &  \mathcal P_{\tilde\sigma\varphi} \\[2ex]
\mathcal P_{\varphi\tilde\sigma}  &   \mathcal P_{\varphi\varphi}
}\pmat{
\displaystyle 0
 &  \mathcal Y^\mu \p^2 \p_\mu \\[3ex]
- \mathcal Y^\nu \p^2 \p_\nu  & \mathcal A^\mu \p_\mu
}\pmat{
\mathcal P_{\tilde\sigma\tilde\sigma} &  \mathcal P_{\tilde\sigma\varphi} \\[2ex]
\mathcal P_{\varphi\tilde\sigma}  &  \mathcal P_{\varphi\varphi}
}
\pmat{
\displaystyle 0
 &  \mathcal Y^\mu \p^2 \p_\mu \\[3ex]
- \mathcal Y^\nu \p^2 \p_\nu  & \mathcal A^\mu \p_\mu
}
\pmat{
\mathcal P_{\tilde\sigma\tilde\sigma} &  \mathcal P_{\tilde\sigma\varphi} \\[2ex]
\mathcal P_{\varphi\tilde\sigma}  &  \mathcal P_{\varphi\varphi}
}
\pmat{
\p_t{\mathcal R}_k^{\tilde\sigma\tilde\sigma} & \p_t{\mathcal R}_k^{\tilde\sigma\varphi}   \\[2ex]
\p_t{\mathcal R}_k^{\varphi\tilde\sigma} &  \p_t{\mathcal R}_k^{\varphi\varphi}
}
}
\Bigg]\,.
}
By extracting the term proportional to $(\p_\mu\bar\phi)^2$, we obtain the flow equation for $K$.

\section{Comparison of flow kernels}
\label{sec:Comparison}

In this appendix, we compare the flow kernels between the present work and Ref.~\cite{Henz:2016aoh} within the large and small field expansions.

\subsection{Small field limit}

The small-field expansion for the flow kernels are
\al{
u(\tilde\rho) &=  \sum_{n=0}^\infty  \frac{u_n}{n!} \tilde\rho^n= u_0 + u_1 \tilde\rho + \cdots\,, \\[1ex]
w(\tilde\rho) &=  \sum_{n=0}^\infty \frac{w_n}{n!} \tilde\rho^n = w_0 + w_1 \tilde\rho + \cdots\,, \\[1ex]
\kappa(\tilde\rho) &=  \sum_{n=0}^\infty \frac{\kappa_n}{n!} \tilde\rho^n = \kappa_0  + \kappa_1\tilde\rho + \cdots\,.
}
In our work we find a UV-fixed point 
\al{
&u_0 = 0.0055\,,&
&w_0 = 0.0277\,,&
&\eta = - \p_t \log \kappa_0  = 0.0116\,,
}
while Ref.~\cite{Henz:2016aoh} finds
\al{
&u_0 = 0.0006460\,,&
&w_0 = 0.002757\,,&
&\eta = - \p_t \log \kappa_0  = 1.6669\,.
}
%

\subsection{Large field limit}

Next we consider the large-field expansion of the flow kernels in terms of $x= 1/(\xi_\infty\tilde\rho)$.
To compare our results with those of Ref.~\cite{Henz:2016aoh}, we adopt the ansatz 
\al{
\label{aeq:uyexpansion}
u(x) &= u_\infty +\sum_{n=0}^\infty  \frac{u_n}{n!} x^n\,, \\[1ex]
\label{aeq:wyexpansion}
w(x) &= x^{-1} + \sum_{n=0}^\infty \frac{t_n}{n!} x^n \,, \\[1ex]
\label{aeq:kyexpansion}
\kappa(x) &= \varepsilon_\infty \xi_\infty + \xi_\infty \sum_{n=1}^\infty  \frac{k_n}{n!} x^n\,,
}
for which we have
\al{
\partial_t u &= -4u - 2x \frac{\partial u}{\partial x}
  + \frac{A_V}{B^3} + \frac{C_V}{B^4} x \,,\\[1ex]
\partial_t w &= -2w - 2x \frac{\partial w}{\partial x}
  + \frac{A_F}{B^3} + \frac{C_F}{B^4} x \,,\\[1ex]
\partial_t \kappa &= -2x \frac{\partial \kappa}{\partial x}
  + \frac{C_K}{B^4} x \,.
}
The coefficients $A_V$, $C_V$ etc. are related to 
\cref{eq:uexpansion,eq:wexpansion,eq:kexpansion} by suitable factors of $B$ and $\xi_\infty$.
Our study gives
\al{
A_V&=\frac{1}{32\pi^2}(3 \epsilon ^3+54 \epsilon ^2+324 \epsilon +648)\,,\nn
C_V&=\frac{1}{480\pi^2}(34 t_0 \epsilon ^4+762 t_0 \epsilon ^3+6372 t_0 \epsilon^2+23544 t_0 \epsilon +32400 t_0 +5 k_1 \epsilon ^3+90 k_1 \epsilon^2+540 k_1 \epsilon +1080 k_1+75 u_\infty \epsilon^4\nn
&\qquad
+1800 u_\infty \epsilon ^3+16200 u_\infty\epsilon^2+64800 u_\infty \epsilon +97200 u_\infty)\,,\nn
A_F&=\frac{1}{1536\pi^2}(509 \epsilon ^3+9356 \epsilon ^2+57300 \epsilon +116928)\,,\nn
C_F&=\frac{1}{322560\pi^2}(37950 t_0 \epsilon ^4+890100 t_0 \epsilon ^3+8070408 t_0 \epsilon ^2+33502896 t_0 \epsilon +53561088 t_0-1680 k_1 \epsilon ^3-71736 k_1 \epsilon ^2\nn
&\qquad
-679392 k_1 \epsilon -1856736 k_1+85505 u_\infty \epsilon^4+2057244 u_\infty \epsilon ^3+18491508 u_\infty \epsilon ^2+73592064 u_\infty \epsilon +109408320 u_\infty)\,,\nn
C_K&=\frac{1}{32\pi^2}(-5 \epsilon ^4-69 \epsilon ^3-360 \epsilon^2-648 \epsilon)\,.
}
Ref.~\cite{Henz:2016aoh} finds
\al{
A_V&=\frac{1}{192\pi^2}( 9\epsilon^3 + 82\epsilon^2 + 612\epsilon + 2760)\,,
\nn
C_V&=\frac{1}{2592\pi^2}( 1080 t_0\epsilon^3 - 3240 t_0 \epsilon^2 -62208 t_0 \epsilon + 1080 k_1 \epsilon^2 - 3240 k_1 \epsilon - 62208 k_1 \nn
&\qquad
+ 383 u_\infty \epsilon^4 + 5004 u_\infty \epsilon^3 + 51120u_\infty\epsilon^2 + 278208 u_\infty \epsilon +  382320 u_\infty)\,,
\nn
A_F&=\frac{1}{3456\pi^2}( -253\epsilon^3 - 6094 \epsilon^2 - 36240\epsilon - 51840)\,,
\nn
C_F&=\frac{1}{5184\pi^2}( 2148t_0 \epsilon^3 - 2916t_0\epsilon^2 - 98712t_0\epsilon + 2310 k_1 \epsilon^2 - 972k_1 \epsilon - 92880 k_1
\nn
&\qquad
- 345 u_\infty \epsilon^4 -23498 u_\infty \epsilon^3 - 213492u_\infty \epsilon^2 - 546552u_\infty\epsilon - 263520 u_\infty)\,,
\nn
C_K&=\frac{1}{36\pi^2}(- \epsilon ^4 + 90 \epsilon^3 + 12636\epsilon^2 - 26244\epsilon)\,.
}

If we use the ansatz \labelcref{eq:uexpansion,eq:wexpansion,eq:kexpansion}, i.e.,
\al{
\p_\tau u(x)&=2 x\p_x u(x)= -4u(x)+ A_u +\frac{C_u}{B}x = 2s_1x +\cdots\,,
\\[1ex]
\p_\tau w(x)&=2 x\p_x w(x)= -2w(x) +\frac{A_w}{B}+\frac{C_w}{B^2}x+\cdots = -\frac{2}{x} + 2t_1 x +\cdots\,,
\\[1ex]
\p_\tau \kappa(x) &=2 x\p_x \kappa(x) = \xi_\infty \frac{C_\kappa}{B}x +\cdots = 2\xi_\infty k_1 x +\cdots\,,
}
the coefficients in Ref.~\cite{Henz:2016aoh} are
\al{
A_u &= 
\frac{1}{93312\pi^2}
\left(
  6210
  - 1728\epsilon
  + 531\epsilon^2
  - 130\epsilon^3
\right)\\
C_u &=
\frac{1}{46656\pi^2}\Big(
-5184 k_1
+ 31860 u_\infty
- 5184 t_0 \epsilon
+ 2322 k_1 \epsilon
+ 7254 u_\infty \epsilon
+ 2322 t_0 \epsilon^2
- 639 k_1 \epsilon^2
- 2022 u_\infty \epsilon^2\nn
&\quad
- 639 t_0 \epsilon^3
+ 150 k_1 \epsilon^3
+ 676 u_\infty \epsilon^3
\Big)\\
A_w &=
\frac{1}{2239488\pi^2}
\Big(
-933120
- 341280\epsilon
+ 29988\epsilon^2
- 5070\epsilon^3
\Big) \\
C_w &= 
\frac{1}{1119744\pi^2}\Big(
-557280 k_1
-1581120 u_\infty
-592272 t_0 \epsilon
+179928 k_1 \epsilon
-2752272 u_\infty \epsilon
+179928 t_0 \epsilon^2
-30636 k_1 \epsilon^2\nn
&\quad
-319608 u_\infty \epsilon^2
-30636 t_0 \epsilon^3
+5214 k_1 \epsilon^3
+42000 u_\infty \epsilon^3
+5214 t_0 \epsilon^4
-887 k_1 \epsilon^4
-7192 u_\infty \epsilon^4
\Big)
\\
C_\kappa &=
\frac{1}{7776\pi^2}
\Big(
-26244\,\epsilon
+ 25758\,\epsilon^2
- 10602\,\epsilon^3
+ 3275\,\epsilon^4
\Big)\,.
}

\section{Fixed point structure and Scaling solutions for \texorpdfstring{$u(\tilde\rho)$}{} with fixed \texorpdfstring{$w(\tilde\rho)$}{}  and \texorpdfstring{$\kappa(\tilde\rho)$}{}}

In this appendix we demonstrate a simple case where we fix $\kappa(\tilde\rho)=1$ and $w(\tilde\rho)=w_0$ (constant).
In this case, the flow equation for $u(\tilde\rho)$ is given by
\al{
\p_t u(\tilde\rho) = - \p_\tau u(\tilde\rho) -4u(\tilde\rho)  -\frac{5 u(\tilde\rho)}{24 \pi ^2 (u(\tilde\rho)-w_0)} -\frac{3u(\tilde\rho)}{160 \pi ^2 (u(\tilde\rho)-4 w_0)} +\frac{79}{480 \pi ^2}\,.
}
The scaling solution thus reads
\al{
 \p_\tau u(\tilde\rho)= -4u(\tilde\rho)  -\frac{5 u(\tilde\rho)}{24 \pi ^2 (u(\tilde\rho)-w_0)} -\frac{3u(\tilde\rho)}{160 \pi ^2 (u(\tilde\rho)-4 w_0)} +\frac{79}{480 \pi ^2}\,.
}
Setting $u(\tilde\rho)=u_0$ and solving
\al{
0 = -4u_0 -\frac{5 u_0}{24 \pi ^2 (u_0-w_0)} -\frac{3u_0}{160 \pi ^2 (u_0-4 w_0)} +\frac{79}{480 \pi ^2}\,.
}

\begin{figure}
    \centering
    \includegraphics[width=0.48\columnwidth]{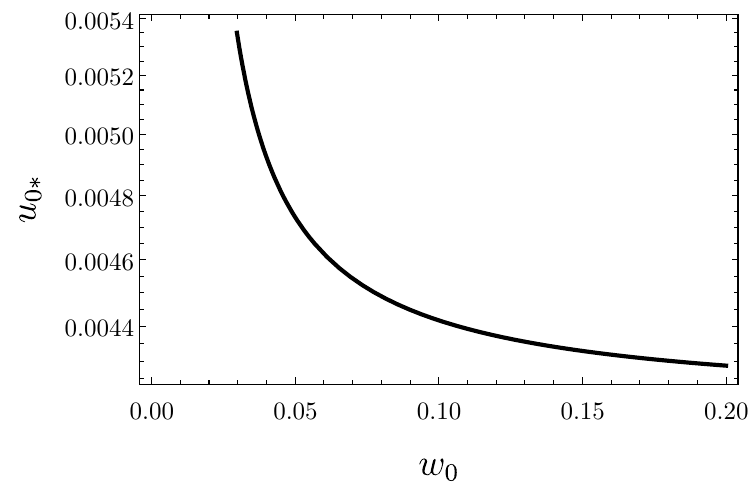}
        \hspace{2ex}
    \includegraphics[width=0.48\columnwidth]{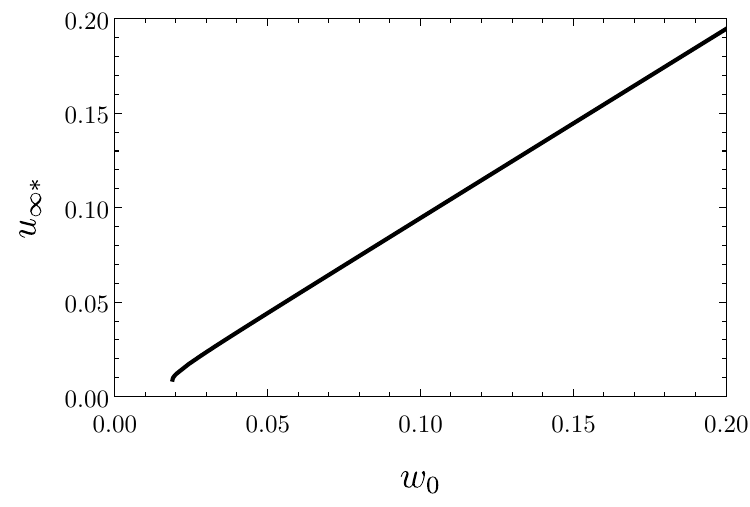}
    \caption{
UV (left) and IR (right) fixed point values of $u_{0}$ and $u_{\infty*}$ varying $w_0$.
    }
    \label{fig: FP u with w0}
\end{figure}

\begin{figure}
    \centering
    \includegraphics[width=0.6\columnwidth]{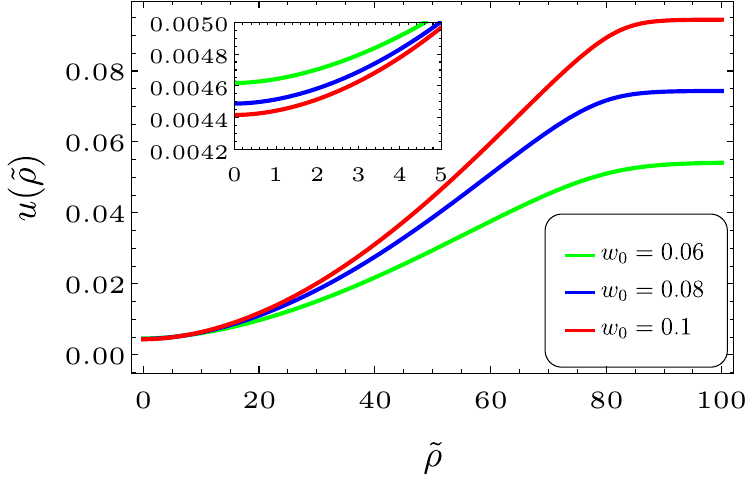}
    \caption{
   Scaling solutions for $u(\tilde\rho)$ with $\kappa(\tilde\rho)=1$ and several fixed values of $w(\tilde\rho)=w_0$.
    }
    \label{fig:u scaling solution}
\end{figure}

One typically finds two fixed points.
One is approached for $\tilde\rho\to \infty$ and denoted by $u_{\infty*}$ (IR-fixed point).
The other is approached for $\tilde\rho\to 0$ and denoted by $u_{0*}$
(UV-fixed point).
For several fixed values of $w_0$, we obtain the IR fixed point $u_{\infty*}$ and the UV fixed point $u_{0*}$, respectively, as
\al{
\text{(IR)}:\quad u_{\infty*}=
\begin{cases}
0.0542678 & \text{for $w_0=0.06$}\,,\\[1ex]
0.0743981 & \text{for $w_0=0.08$}\,,\\[1ex]
0.0944702 & \text{for $w_0=0.1$}\,,
\end{cases} 
}
and
\al{
\text{(UV)}:\quad u_{0*}=
\begin{cases}
0.00461833 & \text{for $w_0=0.06$}\,,\\[1ex]
0.00448945 & \text{for $w_0=0.08$}\,,\\[1ex]
0.00441818 & \text{for $w_0=0.1$}\,.
\end{cases} 
}
The $w_0$ dependence of the IR and UV-fixed point values is depicted in \Cref{fig: FP u with w0}.
The scaling solution for $u(\tilde\rho)$ with several values of $w_0$ is displayed in \Cref{fig:u scaling solution}, illustrating the crossover between the UV and IR fixed points.

\bibliographystyle{JHEP} 
\bibliography{refs}
\end{document}